\documentclass[aps,
final,%
notitlepage,%
twocolumn,
nofootinbib,%
centertags,
floatfix, pra]%
{revtex4-1}
 \pdfoutput=1

\usepackage{color}
\newcommand{\rd}{{\mathrm d}}
\newcommand{\R}{\mathbb R}
\newcommand{\C}{\mathbb C}

\newcommand{\bbone}{{\mathbb I}}

\newcommand{\e}{{\mathrm e}}
\newcommand{\Ccal}{{\mathcal C}}

\newcommand{\Ncal}{{\mathcal N}}
\newcommand{\Pcal}{{\mathcal P}}
\newcommand{\Rcal}{{\mathcal R}}

\newcommand{\rhoG}{\rho^{\mathrm{G}}}
\newcommand{\chiG}{\chi^{\mathrm{G}}}
\newcommand{\WG}{W^{\mathrm G}}

\newcommand{\chiGpm}{\chi^{\mathrm{G}}_{\pm}}
\newcommand{\WGpm}{W^{\mathrm{G}}_{\pm}}
\newcommand{\rhoGm}{\rho^{\textrm G} _-}
\newcommand{\chiGm}{\chi^{\mathrm{G}}_{-}}
\newcommand{\WGm}{W^{\mathrm{G}}_{-}}

\newcommand{\vertiii}[1]{{\left\vert\kern-0.25ex\left\vert\kern-0.25ex\left\vert #1 
		\right\vert\kern-0.25ex\right\vert\kern-0.25ex\right\vert}}
\newcommand{\tR}{t_{\textrm{R}}}

\newcommand{\eq}[1]{\begin{equation} #1 \end{equation}}
\newcommand{\eqstar}[1]{\begin{equation*} #1 \end{equation*}}
\newcommand{\tr}{\mathrm{Tr}}
\newcommand{\Tr}{\mathrm{Tr}}

\newcommand{\nbarinf}{\overline n_\infty}
\newcommand{\eqarray}[1]{\begin{eqnarray} #1 \end{eqnarray}}

\newcommand{\ket}[1]{\vert #1 \rangle}
\newcommand{\bra} [1] {\langle #1 \vert}

\newcommand{\mean}[1]{\langle #1 \rangle}

\newcommand{\cov}[1]{\text{Cov}[#1]}

\newcommand{\QCStwoG}{{\mathcal C}^2_{\mathrm G}}

\newcommand{\cbold}{\boldsymbol{c}}
\newcommand{\mcbold}{\boldsymbol{m}_{\boldsymbol{c}}}
\newcommand{\zbold}{\boldsymbol{z}}

\usepackage{graphicx}
\usepackage{dcolumn}
\usepackage{bm}
\usepackage{color}
\usepackage{soul}
\usepackage{amsmath,amsthm}
\usepackage[dvipsnames]{xcolor}
\usepackage{caption}
\usepackage{subcaption}
\usepackage{mathtools}
\usepackage{dsfont}
\usepackage{amssymb}
\usepackage{enumitem}
\usepackage{ulem}

\usepackage{hyperref}
\hypersetup{
	colorlinks=true,
	linkcolor=ForestGreen,
	filecolor=magenta,      
	urlcolor=cyan,
	citecolor=RoyalBlue
}
\usepackage{pifont}
\usepackage{amssymb}

\usepackage{titlesec}

 \newtheoremstyle{break}
{\topsep}{\topsep}%
{\itshape}{}%
{\bfseries}{}%
{\newline}{}%
\theoremstyle{break}

\newtheorem{lem}{Lemma}

\newtheorem{prop}{Proposition}

\newtheoremstyle{prime}
{\topsep}{\topsep}%
{\itshape}{}%
{\bfseries}{'}%
{\newline}{}%
\theoremstyle{prime}

\begin{document}

\title{Decoherence and nonclassicality of photon-added/subtracted multi-mode Gaussian states}	
	\author{ Anaelle Hertz}
	\affiliation{
		Department of Physics, University of Toronto, Toronto, Ontario M5S 1A7, Canada
	}
\author{Stephan De Bi\`evre}
\affiliation{Univ. Lille, CNRS, Inria, UMR 8524 - Laboratoire Paul Painlev\'e, F-59000 Lille, France}

\begin{abstract}
Photon addition and subtraction render Gaussian states non-Gaussian. We provide a quantitative analysis of the change in nonclassicality produced by these processes by analyzing the Wigner negativity and quadrature coherence scale (QCS) of the resulting states. The QCS is a recently introduced measure of nonclassicality [PRL 122, 080402 (2019), PRL 124, 090402 (2020)], that we show to undergo a relative increase under photon addition/subtraction that can be as large as 200\%.  This implies that the degaussification and the concomitant increase of nonclassicality  come at a cost. Indeed, the QCS  is  proportional to the decoherence rate of the state so that the resulting states are considerably more prone to environmental decoherence. Our results are quantitative and rely on explicit and general expressions for the characteristic and Wigner functions of photon added/subtracted single- and multi-mode Gaussian states for which we provide a simple and straightforward derivation. These expressions further allow us to certify the quantum non-Gaussianity of the photon-subtracted states with positive Wigner function.
\end{abstract}

\maketitle

\section{Introduction}
Gaussian states are prominent in continuous-variable quantum information as they are relatively easy to produce experimentally  and simple to study theoretically. Nevertheless, non-Gaussian states or operations are essential for performing certain quantum information tasks. They are for example needed to achieve universal photonic quantum computation\cite{Lloyd,Pant}.
One possible method for producing non-Gaussian states is through  photon addition or subtraction from a Gaussian state. This technique has attracted interest because it allows the engineering of a variety of non-Gaussian quantum states. It has for example been shown that cat states with small amplitude can be prepared with a fidelity close to one by subtracting a photon from a vacuum squeezed state~\cite{Zavatta,Ourjoumtsev,Neergaard}. Over the last few years, a variety of experimental techniques have been developed to generate and study photon-added/subtracted Gaussian states \cite{Zavatta,Wenger,Parigi,Parigi2007, Kiesel, Biagi2020, Ra2020, Biagi2021}. 
For reviews on photon addition and subtraction, we refer to~\cite{KimReview, Walschaers21}.

For a state to be non-Gaussian is however not always enough for it to be interesting in the context of quantum information or quantum computing tasks. Indeed, non-Gaussian states may still be classical meaning that they may be mixtures of coherent states. Or they may be more generally mixtures of Gaussian states, in which case they are said not to be quantum non-Gaussian, or genuinely non-Gaussian.  (See~\cite{Genoni13, Takagi18, LaStraHloJezFi19, Walschaers21} and references therein for details on the latter subject.) Nonclassicality or the stronger property of quantum (or genuine) non-Gaussianity are needed for certain quantum informational tasks and a variety of techniques for their detection and measure have been developed.  In this paper, we provide a quantitative analysis of the degree to which photon-added/subtracted Gaussian states are nonclassical or quantum non-Gaussian.

For our analysis, we will concentrate on two distinct measures of nonclassicality/non-Gaussianity, namely  their quadrature coherence scale (QCS) and their Wigner negativity, as expressed through their Wigner negative volume. The QCS is a recently introduced nonclassicality measure ~\cite{Debievre, Hertz}, the definition and main nonclassical features of which are recalled in Sec.~\ref{s:QCS}.   The Wigner negativity, on the other hand, is a common measure of nonclassicality and has been shown to be a monotone in a resource theory of quantum non-Gaussianity~\cite{Takagi18}. 

Our results for single-mode states, detailed below, establish that  the degaussification through photon addition/subtraction does substantially enhance the nonclassical features of the underlying Gaussian states. At low and intermediate squeezing, photon addition is more efficient in doing so, but at high enough squeezing, photon addition or subtraction are shown to be equivalent in this respect. Importantly, these results also entail that the increased nonclassicality that is generated in the process comes at a cost. Indeed, the QCS of a state is proportional to its decoherence rate~\cite{Hertz}, so that a large value of the QCS is equivalent to a short decoherence time. The photon-added/subtracted states are therefore much more sensitive to environmental decoherence than their Gaussian mother states. And the photon-added states tend to be considerably more sensitive than the photon-subtracted ones. 

More precisely, we show that the Wigner negative volume of single-mode photon-added/subtracted Gaussian states  reaches its maximal value when there is no noise, hence on the photon-added/subtracted squeezed vacuum states. This maximum is independent of the amount of squeezing.
In the presence of noise, and at low to intermediate values of the squeezing, we show the Wigner negative volume is more sensitive to noise and hence 
smaller  for photon-subtracted squeezed Gaussian states  than for photon-added ones. This means that, at intermediate squeezing, the intuitive idea that photon-addition is more efficient than photon-subtraction in producing nonclassical features such as Wigner negativity  \and hence quantum non-Gaussianity is indeed correct for Gaussian states.  One should note however that, as we show,  there is a tradeoff between squeezing and Wigner negative volume: photon-added squeezed thermal states loose Wigner negative volume as the squeezing is increased. At large squeezing, the advantage of photon-addition over photon-subtraction is diminished:  we establish that the Wigner negative volume  is then identical for photon-added and photon-subtracted states. 

Concerning the QCS of photon-added/subtracted single-mode Gaussian states, we show that the degaussification accomplished by photon addition/subtraction does typically increase the QCS, and hence the associated nonclassical features of the state, and that this increase is often substantial. As for the Wigner negative volume, it is more pronounced for photon-addition than for photon-subtraction, except at large values of the squeezing, where it is again asymptotically identical.

As a byproduct of our analysis, we show a number of structural results about photon addition/subtraction that are of independent interest and valid for arbitrary $n$-mode Gaussian states $\rhoG$. While photon addition/subtraction is guaranteed to make $n$-mode Gaussian states non-Gaussian, it is known, and very easy to see, that photon subtraction applied to a $n$-mode {\it classical} Gaussian state yields a {\it classical} non-Gaussian state. We will show that, in addition, photon-subtraction, applied to a $n$-mode  {\it nonclassical} Gaussian state yields a {\it nonclassical} non-Gaussian state (Proposition~\ref{prop:rhominusclass}). Note that this is not true for non-Gaussian states: for example, the one-photon Fock state, which is nonclassical, is transformed into the vacuum, which is classical. In addition, we show that,
if a single-mode photon-subtracted Gaussian state $\rhoGm\sim a(\boldsymbol c)\rhoG a^\dagger(\boldsymbol c)$ (see Eq.~\eqref{eq:rhominusdef}) is Wigner negative, then the underlying Gaussian state $\rhoG$ has a QCS strictly larger than $1$ 
(Lemma~\ref{Lemma:WigNegSqThmoins1mode}).  
This is in contrast to what happens with photon addition, which, applied to any Gaussian state, is known~\cite{Fabre, FabrePRL17} to always produce a Wigner negative and hence nonclassical state. 
We further use  a sufficient criterium for quantum non-Gaussianity in terms of the Wigner function from~\cite{Genoni13}  to identify a family of photon-subtracted Gaussian state with positive Wigner function that are quantum non-Gaussian.

The paper is organized as follows.  In Sec. \ref{PhaseSpaceFormalism} we  give a brief review of the phase space formalism of quantum optics. In Sec.~\ref{s:QCS} we introduce the QCS and recall its main features as a nonclassicality witness and measure. To compute it for photon-added/subtracted states, we need their Wigner and/or characteristic functions. We show how to straightforwardly compute those for general multi-mode photon-added/subtracted states in Sec.~\ref{ChiPhotonAdded} and apply the result when the initial state is Gaussian. The resulting formulas are simply expressed in terms of the covariance matrix and displacement operator characterizing the initial Gaussian state: see Eq.~\eqref{eq:CharacPhotonAddedGaussianState} and Eq.~\eqref{eq:Wignerpm}. Equivalent, but less explicit formulas were obtained previously in~\cite{Walschaers21, Fabre, FabrePRL17}, using a more complex and considerably more lengthy derivation.
We first use these expressions to make  a number of general qualitative and quantitative observations on the (non)classicality and Wigner negativity/positivity of photon-subtracted multi-mode Gaussian states in Sec.~\ref{s:nonclWN}.   
In Sec. \ref{s:SqThPlus}-\ref{s:SqThminus} we then turn to a quantitative study of the Wigner negative volume and of the relative change in the QCS for single-mode photon-added/subtracted squeezed thermal states. 
In Sec.~ \ref{s:2mode} we discuss the two-mode case through some illustrative examples, some of which have recently been prepared experimentally~\cite{Biagi2020}. 
We conclude and discuss some open problems  in Sec. \ref{Conclusion}.


\section{Phase space formalism}
\label{PhaseSpaceFormalism}
	We start by briefly introducing  the symplectic formalism employed for continuous-variable states in quantum optics. More details can be found, for example, in \cite{weedbrook, Anaellethesis}.
	
	A continuous-variable system is represented by $n$ modes. To  each of them are associated the annihilation/creation operators $a_i$ and $a_i^\dag$ verifying the commutation relation $[a_i,a_i^\dag]=1$. We define the vector of quadratures $\boldsymbol{\hat r}=(\hat{x}_1,\hat{p}_1,\hat{x}_2,\hat{p}_2,\cdots,\hat{x}_n,\hat{p}_n)$  where
	\begin{equation*}
	\hat{x}_j=\frac1{\sqrt{2}}(a_j+a_j^\dag),\quad \hat{p}_j=-\frac{i}{\sqrt{2}}(a_j-a_j^\dag)\quad \forall j=1,\cdots,n.
	\end{equation*}
	Each quantum state $\rho$ can be described by a characteristic function 
\eq{\chi(\boldsymbol{z})=\tr \rho D(\boldsymbol{z})\label{eq:charfunct}}
 where 
 	\eq{D(\boldsymbol z)=\exp\left\{\sum_{j=1}^n(z_ja_j^\dag-\bar z_j a_j)\right\}=e^{a^\dag(\boldsymbol z)- a(\boldsymbol{z})}\label{displacementOperator}}
 		is the n-mode displacement operator and where
 	\eq{a^\dag(\boldsymbol z)=\sum_i z_i a^\dag_i,\qquad\qquad a(\boldsymbol z)=\sum_i\bar z_i a_i.\label{defadag}}
Note that, for any $\boldsymbol{z}, \boldsymbol{z}'\in\C^n$,
$$
[a(\boldsymbol{z}), a^\dagger(\boldsymbol{z}')]=\sum_i \overline z_i z'_i=\overline{\boldsymbol{z}}\cdot \boldsymbol{z}'.
$$
For later use, we recall
\begin{eqnarray}
a^\dagger(\boldsymbol{c})D(\boldsymbol{z})&=&D(\boldsymbol{z})(a^\dagger(\boldsymbol{c})+\overline{\boldsymbol{z}}\cdot \boldsymbol{c})\label{eq:pullthrough1}\\
a(\boldsymbol{c})D(\boldsymbol{z})&=&D(\boldsymbol{z})(a(\boldsymbol{c})+{\boldsymbol{z}}\cdot \overline{\boldsymbol{c}}). \label{eq:pullthrough2}
\end{eqnarray}

	 The Fourier transform of the characteristic function gives the Wigner function 
	\eq{W(\boldsymbol{\alpha})=\frac{1}{(\pi^2)^n}\int\chi(\boldsymbol{z})e^{(\boldsymbol{\bar z\cdot\alpha}-\boldsymbol{z\cdot\bar{\alpha}})}d^{2n}\boldsymbol{ z}\label{FourierWigner}}
where $d^{2n} \boldsymbol z=d^n\text{Re}(\boldsymbol z)d^n\text{Im}( \boldsymbol z)$,	$\boldsymbol\alpha=\begin{pmatrix}\alpha_1&\cdots&\alpha_n\end{pmatrix}^T$ and $\alpha_j=\alpha_{j1}+i\alpha_{j2}=\frac1{\sqrt{2}} (x_j+ip_j)\in\C$. 
	It is normalized so that $\int W(\boldsymbol\alpha) d^{2n}\boldsymbol{\alpha}=\chi(0)=1$.
	
For later reference, we recall that a state $\rho$ is said to be optically classical~\cite{Titulaer} if and only if there exist a positive function $P(\boldsymbol{z})$ so that
\begin{equation}\label{eq:class-state}
\rho=\int P(\boldsymbol{z}) |\boldsymbol{z}\rangle\langle \boldsymbol{z}| \rd \boldsymbol{z}.
\end{equation}
Here $|\boldsymbol{z}\rangle=D(\boldsymbol{z})|0\rangle$ are the coherent states with $|0\rangle$ the vacuum state. Otherwise, the state is said to be optically nonclassical. In other words, a state is said to be optically nonclassical if it is not a mixture of coherent states. In what follows, we will drop ``optically'' from ``optically nonclassical''.

	The first-order moments of a state $\rho$ constitute the displacement vector, defined as $\boldsymbol{d}=\mean{\boldsymbol{\hat r}}=\tr (\boldsymbol{\hat r}\rho)$, while the second moments make up the covariance matrix $V$ whose elements are given by
	\begin{equation}
	V_{ij}=2\cov{\hat r_i,\hat r_j}=\mean{\{\hat r_i,\hat r_j\}}-2\mean{\hat r_i}\mean{\hat r_j}
\label{covmat}	\end{equation}
	where $\{\cdot,\cdot\}$ represents the anticommutator.  
	
	A \textit{Gaussian} state $\rhoG$ is fully characterized by its displacement vector and covariance matrix. Its characteristic function is a Gaussian:
	\eqarray{\label{chiGaussian}\chiG(\boldsymbol{\xi})&=&e^{-\frac12\boldsymbol{\xi}^T\Omega V \Omega^T\boldsymbol{\xi}-i \sqrt{2}(\Omega \boldsymbol{d})^T\boldsymbol{\xi}},}
with
$$
\Omega=\bigoplus_{j=1}^n\begin{pmatrix}0&1\\-1&0	\end{pmatrix}.
$$
Here, for all $1\leq i\leq n$, $\xi_i^T=(\xi_{i1}, \xi_{i2})\in\R^2$ and $\boldsymbol{\xi}^T=~(\xi_1^T,\dots, \xi_n^T)\in\R^{2n}$. Also, we define
\begin{equation}\label{eq:z}
z_j=\xi_{j1}+i\xi_{j2},
\end{equation}
 and $\boldsymbol{z}=(z_1,\dots, z_n)\in\C^n$ and we will write, with the  usual abuse of notation: $\chiG(\boldsymbol{ z})=\chiG(\boldsymbol{\xi})$. 

The Wigner function $W^{\textrm{G}}(\boldsymbol{\alpha})$ of a Gaussian state is also a Gaussian. See~\ref{Appendix D} for the explicit expression.


\section{The quadrature coherence scale}\label{s:QCS}
The quadrature coherence scale $\Ccal(\rho)$ (QCS) of a state $\rho$ is defined 
as~\cite{Debievre, Hertz}
\eq{\Ccal^2(\rho)=\frac1{2n\mathcal{P}}\left(\sum_{j=1}^{2n}\tr[\rho,\hat r_j][\hat r_j,\rho]\right)\label{eq:QCSdef} }
where   $\mathcal{P}=\tr\rho^2$ is the purity of the state $\rho$. A summary of its main features is given in this Section.

The expression Eq.~\eqref{eq:QCSdef} for $\Ccal(\rho)$ does not explain why it is called the quadrature coherence scale. To see this, we consider for simplicity of notation the case where only one mode is present: the general case is obtained by taking an average over the modes. It turns out that the QCS can be rewritten as follows:
\begin{eqnarray}\label{eq:QCS2}
\Ccal^2(\rho)&=&\frac1{2\Pcal}\left(\int (x-x')^2 |\rho(x,x')|^2\rd x\rd x'\right. \nonumber\\
&\ &\hskip0.3cm+\left.\int (p-p')^2|\rho(p,p')|^2\rd p\rd p' \right).
\end{eqnarray}
Here $\rho(x,x')$ (respectively $\rho(p,p')$) is the operator kernel of $\rho$ in the $\hat x$-representation (respectively $\hat p$-representation). Since $|\rho(x,x')|^2/\Pcal$ (respectively $|\rho(p,p')|^2/\Pcal$) is a probability distribution, one readily sees the first (second) term in this expression provides the  scale (squared) on which the coherences, meaning the off-diagonal matrix elements $\rho(x,x')$ (respectively $\rho(p,p')$) of the density matrix $\rho$ live. Roughly speaking, one can think of $\rho(x,x')$ and $\rho(p,p')$ as matrices; the square root of the first (respectively second) term in Eq.~\eqref{eq:QCS2} provides the width of the strip parallel to its diagonal in which the $\hat x$-coherences (respectively $\hat p$-coherences) of $\rho$ are substantial. 
It follows that a large $\Ccal(\rho)$ implies that either the $\hat x$- or $\hat p$-coherences live far from the diagonal.  Conversely, a small $\Ccal(\rho)$ implies that the off-diagonal coherences of both quadratures must be small away from the diagonal. As explained in~\cite{Hertz}, a large value of the QCS manifests itself in nonclassical phenomena such as fast oscillations of the Wigner function, of the probability densities $\rho(x,x)$ and/or $\rho(p,p)$ for position and momentum and of the photon number probability, which can be interpreted as interference phenomena.
 
In fact, as pointed out already in the introduction, $\Ccal^2(\rho)$ provides  a measure of optical non-classicality. More precisely, $\Ccal^2(\rho)>1$ implies $\rho$ is nonclassical and a large value of the QCS corresponds to a large nonclassicality~\cite{Debievre}. Coherent states, on the other hand, have a QCS equal to $1$; all other classical states have a QCS less than or equal to $1$, which is therefore a natural reference value for the QCS.  The evaluation of the QCS on large families of benchmark states in~\cite{Debievre, Hertz, Horoshko, HertzCerfDebievre} confirms the efficiency of the QCS as an optical nonclassicality measure.  For example, highly excited Fock states, cat states with large separation, highly squeezed states and strongly entangled states all have a large QCS. Some explicit examples of this type are given below in this section. In the following sections, the QCS of photon-added/subtracted Gaussian states  will be studied in detail and the results will confirm this general picture. 

The QCS has recently be shown to be experimentally accessible. In~\cite{Griffetetal22} an interferometric scheme was proposed allowing a direct measurement of the QCS using two identical copies of the state, thereby avoiding having recourse to a full state tomography. This scheme has then been carried out on a cloud quantum computer~\cite{Goldbergetal23}.

For our purposes here, a second feature of the QCS is crucial. It was proven in~\cite{Hertz} that the QCS is directly related to the decoherence time of $\rho$, as follows. When coupled to a thermal bath, and provided $\Ccal^2(\rho)>1$, the half life $\tau_\Pcal$ of the purity of $\rho$ satisfies 
$$
\tau_\Pcal\approx\frac12\frac1{(2\nbarinf+1)\Ccal^2(\rho)-1}\tR,
$$
where $\tR$ is the time scale on which the system converges to the thermal equilibrium with mean photon number $\bar n_\infty$, which characterizes the temperature of the bath. Similarly, the half life of $\Ccal^2(\rho)$ itself is also inversely proportional to $\nbarinf \Ccal^2(\rho)$. In other words, the speed at which environmental decoherence takes place is proportional to the QCS (squared). 

In conclusion, whereas a large QCS does imply strong nonclassical properties of the state, as recalled above, this nonclassicality is accompanied automatically with an increased sensitivity to environmental decoherence and hence to a shorter decoherence time. For more details we refer to~\cite{Hertz}. 

In what follows, we investigate how the QCS of Gaussian states is affected by photon addition or subtraction. This will inform us on the change in decoherence time of the degaussified states, compared to the original Gaussian state. We will see that, as a rule, the degaussified state has a much larger QCS, hence a much shorter decoherence time.

For our purposes, neither the expression in Eq.~\eqref{eq:QCSdef} nor the one in~Eq.~\eqref{eq:QCS2} are suitable. It is shown in~\cite{Debievre} that the QCS for a general $n$-mode state can be written in terms of the Wigner or characteristic function of the state:
\begin{equation}\label{eq:QCSWch}
\Ccal^2(\rho)=\frac{\||\xi|\chi\|_2^2}{n\|\chi\|_2^2}=\frac14\frac{\|\nabla W\|_2^2}{n\|W\|_2^2}.
\end{equation}
Here, with $\xi,\alpha\in \C^n$ and $\|\cdot\|_2$ stands for the $L^2$-norm, meaning for example $\|W\|_2^2:=\int|W(\alpha)|^2\rd^{2n} \alpha$. 
The expressions obtained for the Wigner and characteristic function of photon-added/subtracted states in the next section will allow us to compute their QCS and the corresponding change in QCS. 

Let us note that, for pure states, a simple computation starting from \eqref{eq:QCSdef} shows that 
\begin{equation}\label{eq:pureQCS}
\Ccal^2(\rho)=\frac1{n}\sum_i(\Delta \hat x_i)^2 +(\Delta \hat p_i)^2,
\end{equation}
 which is the so-called total noise of $\rho$~\cite{sc86}. As a result, for the $n$-th Fock state $\ket{n}$, one finds
 \begin{equation}\label{eq:FockQCS}
 \Ccal^2(|n\rangle\langle n|)=2n+1
\end{equation}
and for cat states $|\psi_\pm\rangle\simeq|\alpha\rangle\pm|-\alpha\rangle$, one has $\Ccal^2(|\psi_\pm\rangle\langle\psi_\pm|)\simeq 2|\alpha|^2$. 

For an $n$-mode Gaussian state $\rhoG$, pure or mixed, one finds~\cite{Hertz, HertzCerfDebievre}
 \begin{equation}\label{eq:QCSGaussian}
 \QCStwoG=\Ccal^2(\rhoG)=\frac1{2n}\Tr V^{-1}.
 \end{equation}
 For example, the 
 squeezed thermal states, defined in Eq.~\eqref{eq:SqTh},  have $\Ccal^2_{\textrm{SqTh}}=\frac{1-q}{1+q}\cosh r$ (see Eq.~\eqref{eq:QCSSqTh}). Note the growth of the QCS with $n$, $\alpha$ and the squeezing parameter $r$ respectively. 
 
 We will continue the practice of~\cite{Debievre, Hertz} in  referring  to optically nonclassical states $\rho$ for which the QCS is less than $1$ as weakly nonclassical states, the others being strongly nonclassical. In other words, we have that 
 $$
 \Ccal^2(\rho)\leq 1
 $$
 if and only if $\rho$ is either classical or weakly nonclassical. The relevance of this boundary between weakly and strongly nonclassical is clear from the many benchmark states investigated previously, and will emerge again below  in Sec.~\ref{s:nonclWN} and in Sec.~\ref{s:SqThminus}.

\section{Characteristic and Wigner functions of multi-mode photon-added/subtracted states}
\label{ChiPhotonAdded}

\subsection{General photon-added/subtracted states}\label{ChiPhotonAddedgen}

We first define what we mean by a general photon-added $n$-mode state $\rho_+$. Recall that the most general multi-mode one-photon state is of the form
\eqstar{\ket{\boldsymbol{c}}=a^\dag(\boldsymbol{c})\ket{0},} 
where $a^\dag(\boldsymbol{c})$ is given by Eq. \eqref{defadag},  $\boldsymbol c\in \C^n$ and $\sum_i|c_i|^2=~1$. 
In general, a photon-added state is  then defined as
\eq{\rho_+=\mathcal{N}_+\, a^\dag(\boldsymbol c)\rho\, a(\boldsymbol c)\
\text{with}
\ \mathcal{N}_+=\left(\tr \left[a^\dag(\boldsymbol c)\rho\,a(\boldsymbol c)\right]\right)^{-1},}
where $\rho$ is the initial or mother state to which a photon is added. Similarly, the photon-subtracted state is defined as
\eq{\rho_-=\mathcal{N}_-\, a(\boldsymbol c)\rho\, a^\dag(\boldsymbol c)\ \text{with}\ \mathcal{N}_-=\left(\tr \left[a(\boldsymbol c)\rho\,a^\dag(\boldsymbol c)\right]\right)^{-1}.\label{eq:rhominusdef}} 
Note that
$$
\tr \left[a^\dag(\boldsymbol c)\rho\,a(\boldsymbol c)\right] = \tr \left[a(\boldsymbol c)\rho\,a^\dag(\boldsymbol c)\right]+1\geq 1,
$$
so that $0<\Ncal_+\leq 1$. However, $\tr \left[a(\boldsymbol c)\rho\,a^\dag(\boldsymbol c)\right]$ can vanish, in which case $a(\boldsymbol{c})\rho a^\dagger(\boldsymbol{c})=0$ so that $\rho_-$ is not defined. We will come back to this point below, but for now we assume $\Ncal_-<+\infty$.
 
We write $\chi_\pm$ for the characteristic function of $\rho_\pm$. Its expression is obtained by a short and straightforward computation and we find:
\eqarray{\label{chiPlus}\chi_\pm(\boldsymbol z)&=&-\mathcal{N}_\pm\left[\boldsymbol{c}\cdot\left(\partial_{ \boldsymbol z}\mp\frac{\bar{\boldsymbol{ z}}}{2}\right)\right]\left[\bar{\boldsymbol{c}}\cdot\left(\partial_{ \bar{\boldsymbol z}}\mp\frac{\boldsymbol{ z}}{2}\right)\right]\chi(\boldsymbol z)\nonumber\\
}
where $\chi(\boldsymbol z)$ is the characteristic function of the state $\rho$. To see this, we note first that
 the displacement operator can be written as
$$
D(\boldsymbol z)=\e^{a^\dagger(\boldsymbol z)}\e^{-a(\boldsymbol z)}\e^{-|\boldsymbol z|^2/2}
$$or equivalently, as
$$D(\boldsymbol z)=\e^{-a(\boldsymbol z)}\e^{a^\dagger(\boldsymbol z)}\e^{|\boldsymbol z|^2/2}.
$$
Consequently, one has the well known formulas
\begin{eqnarray*}
	\partial_{z_j} D(\boldsymbol z)=&\left(a^\dagger_j-\frac{\overline z_j}{2}\right)D(\boldsymbol z)= D(\boldsymbol z)\left(a^\dagger_j+\frac{\overline z_j}{2}\right),\\
\partial_{\overline z_j} D(\boldsymbol z)=&-D(\boldsymbol z)\left(a_j+\frac{z_j}{2}\right)=-\left(a_j-\frac{z_j}{2}\right)D(\boldsymbol z).
\end{eqnarray*}
Hence, for all $\boldsymbol c\in\C^n$, a short computation shows that
$$
-\left[\overline{\boldsymbol c}\cdot\partial_{\bar{\boldsymbol z}}-\frac{\overline{\boldsymbol c}\cdot \boldsymbol z}{2}\right]\left[\boldsymbol c\cdot\partial_{\boldsymbol z}-\frac{\boldsymbol c\cdot\overline{\boldsymbol z}}{2}\right] D(\boldsymbol z)= a(\boldsymbol c)D(\boldsymbol z)a^\dagger(\boldsymbol c).
$$
This implies~Eq.~\eqref{chiPlus} for $\chi_+$. The proof for $\chi_-$ is similar.

 It is clear from Eq. \eqref{chiPlus}  that, when adding  $m$ photons,  one needs to apply $m$ times the operator $-\left[\boldsymbol{c}\cdot\left(\partial_{ \boldsymbol z}-\frac{\bar{\boldsymbol{ z}}}{2}\right)\right]\left[\bar{\boldsymbol{c}}\cdot\left(\partial_{ \bar{\boldsymbol z}}-\frac{\boldsymbol{ z}}{2}\right)\right]$ and to normalize the result.

To compute the Wigner function $W_\pm(\boldsymbol{\alpha})$ of $\rho_\pm$ it now suffices to compute the Fourier transform of $\chi_\pm(\boldsymbol{ z})$ [see Eq. \eqref{FourierWigner}]. Details of the calculation can be found in  \ref{Appendix B}. We obtain
	\eqarray{	\label{WignerPlus}W_\pm(\boldsymbol{\alpha})&=&
\mathcal{N}_\pm\left[\boldsymbol{c}\cdot\left(\frac{\partial_{ \boldsymbol\alpha}}{2}\mp\bar{\boldsymbol{\alpha}}\right)\right]\left[\bar{\boldsymbol{c}}\cdot\left(\frac{\partial_{ \bar{\boldsymbol\alpha}}}{2}\mp\boldsymbol{\alpha}\right)\right]W(\boldsymbol\alpha).\nonumber\\
}
The one-mode version of this expression was already obtained in \cite{Braun}.

Clearly then, if the characteristic function $\chi$ (or Wigner function W) of $\rho$ is known, the characteristic/Wigner function of an arbitrary photon-added/subtracted state can be straightforwardly computed. We illustrate this in the following paragraph for Gaussian states. 

\subsection{Photon-added/subtracted Gaussian states}\label{gaussiancase}
We suppose now that $\rho=\rhoG$ is Gaussian. The computation in  Eq. \eqref{chiPlus} then reduces to elementary algebra, using~\eqref{chiGaussian}. The details are given in  \ref{Appendix C} and the result is
 	\eq{\label{eq:CharacPhotonAddedGaussianState}\chiGpm(\boldsymbol z)
 		=\mathcal{N}_\pm \left(\frac12
 		\overline{\boldsymbol{m}}_{\boldsymbol{c}}^TV\boldsymbol{m}_{\boldsymbol{c}}\pm\frac12
 		-\boldsymbol{ \beta}^T_\pm\boldsymbol{m}_{\boldsymbol{c}}\overline{\boldsymbol{m}}_{\boldsymbol{c}}^T\boldsymbol{ \beta}_\pm
 		\right)\chiG(\boldsymbol z).}
 Here the covariance matrix $V$ is the one of the  Gaussian mother state,
 \begin{equation}\label{eq:betapm}
 \boldsymbol{ \beta}_\pm=\frac12\left(\Omega V \Omega\mp\bbone\right) U^\dag\boldsymbol{ Z}+i\Omega\boldsymbol{ d},
 \end{equation}
  the matrix $U$ is given by \cite{Serafini}
 	\eqstar{\label{eq:udef}U=\bigoplus_{j=1}^n u\qquad \text{where}\qquad
 		u=\frac1{\sqrt{2}}\begin{pmatrix}1&i\\1&-i\end{pmatrix},
 	} and 
 	\eq{\boldsymbol{Z}=\begin{pmatrix} z_1\\ \overline z_1\\\vdots \\ z_n\\ \overline{z}_n \end{pmatrix}=\sqrt2U\boldsymbol \xi,\quad\text{}\qquad\boldsymbol{m}_{\boldsymbol{c}}= U^\dag\begin{pmatrix}
 		c_1\\0\\\vdots\\c_n\\0
 		\end{pmatrix}.\label{eq:mc}
 	}

Note that 
$$
\overline{\boldsymbol m}_{\boldsymbol c}^TV\boldsymbol{m}_{\boldsymbol{c}}= 2\text{Cov}[a^\dag(\boldsymbol{ c}),a(\boldsymbol{ c})].
$$

With analogous calculations (see \ref{Appendix D}), one also finds the Wigner function of a photon-added/subtracted Gaussian state. The resulting expressions are similar with the difference that they involve 
the inverse of the covariance matrix $V$.
One finds \footnote{ With the  usual abuse of notation, we write: $W(\boldsymbol{ \alpha})=W(\boldsymbol{r})$. }
\eq{\WGpm(\boldsymbol r)
	=\mathcal{N}_\pm\Bigg(M_\pm(V,\boldsymbol{c})+	\boldsymbol{ \lambda}^T_\pm
	\boldsymbol{m}_{\boldsymbol{c}}  \overline{\boldsymbol{m}}_{\boldsymbol{c}}^T
	\boldsymbol{ \lambda}_\pm
\Bigg)\WG(\boldsymbol r)
\label{eq:Wignerpm}
}
where 
\begin{equation}\label{eq:lambdapm}
\boldsymbol{ \lambda}_\pm=\left[\left(V^{-1}\pm\bbone\right)\boldsymbol{ r}-V^{-1}\boldsymbol{ d}\right]\in\R^{2n},
\end{equation} 
 and 
 $M_\pm(V,\boldsymbol{c})${$\in\R$} is independent of $\boldsymbol{r}$ and given by
\begin{equation}\label{eq:KpmAc}
M_\pm(V,\boldsymbol{c})=\mp\frac12-\frac12 \overline{\boldsymbol{m}}_{\boldsymbol{c}}^T V^{-1} \boldsymbol{m}_{\boldsymbol{c}}.
\end{equation}
Let us note that in~\cite{ Fabre,FabrePRL17} expressions for the characteristic and Wigner functions of photon-added/subtracted Gaussian states were derived through a rather involved computation of the truncated correlation functions of the states, which then need to be summed. The resulting expressions are however less directly formulated in terms of the covariance matrix $V$ and displacement vector $\boldsymbol{d}$ characterizing the Gaussian mother state. Our derivation here, starting as it does from the general and straightforward expressions in Eq.~\eqref{chiPlus} and~\eqref{WignerPlus}, is elementary and the results are  simply expressed in terms of $\boldsymbol{d}$ and of the (inverse of) $V$. We use them now to analyze the QCS and Wigner negativity of the photon-added/subtracted Gaussian states. 
Let us mention that yet another approach to the computation of the Wigner function of photon-subtracted states is proposed in~\cite{Walschaers21}; the resulting expressions are again less explicit than the ones proposed here. 

\section{(Non)Classicality and Wigner negativity of photon-subtracted Gaussian states.}\label{s:nonclWN}
To prepare our  quantitative analysis of the QCS of photon-added/subtracted Gaussian states, we obtain in this section general results on the (non)classicality and Wigner negativity/positivity of photon-subtracted Gaussian states. We know that photon-addition/subtraction degaussifies any Gaussian state. The question we address is: under what conditions on the Gaussian state and on $\boldsymbol c$ does it become nonclassical or even Wigner negative? Note that, for photon-addition, the answer is immediate. Photon-addition transforms any Gaussian state, centered or not, classical or not, into a Wigner negative and hence nonclassical  and even quantum non-Gaussian state. This follows directly from Eq.~\eqref{eq:Wignerpm}-\eqref{eq:KpmAc} and was pointed out already in~\cite{Fabre, FabrePRL17}. We therefore concentrate on the photon-subtracted case.

For one-mode photon-subtracted Gaussian states, we establish a relation between Wigner negativity and the QCS. Recall that a state is said to be Wigner positive if its Wigner function is everywhere nonnegative. Otherwise it is said, somewhat abusively, to be Wigner negative.  

\subsection{(Non)Classicality of photon-subtracted Gaussian states.}\label{s:noncl}
It is well known that photon-subtraction transforms a classical state into a classical state. We recall the argument. Suppose $\rho$ is classical and let $P(\boldsymbol{z})$ be its $P$-function, which is nonnegative. Then it follows directly from Eq.~\eqref{eq:class-state} that 
$\Ncal_-|\overline{\boldsymbol{c}}\cdot \boldsymbol{z}|^2 P(\boldsymbol z)$, which is still nonnegative, 
is the $P$-function of $\rho_-$. In addition, photon subtraction can make a nonclassical state classical: $a_1|1\rangle=|0\rangle$ is an example. In other words, while photon subtraction always preserves the classicality of states, it does not always preserve their nonclassicality. 

We show here that, nevertheless, photon subtraction always transforms a \emph{Gaussian} nonclassical state into a nonclassical state.  In other words, photon-subtraction preserves both the classicality and the nonclassicality of \emph{Gaussian} states. This is the content of Proposition~\ref{prop:rhominusclass} below.  It generalizes an observation made in  \cite{Biswas} where it is remarked that, under photon subtraction, a single-mode squeezed vacuum state remains nonclassical for all values of squeezing $r>0$. Our result holds for all nonclassical Gaussian multi-mode states, centered or not.

As a preliminary step, we first identify those $\boldsymbol{c}\in\C^n$ with $\overline\cbold\cdot\cbold=1$, and $\rhoG$ for which $a(\boldsymbol{c})\rhoG a^\dagger(\boldsymbol{c})=0$;  for such $\cbold$ and $\rhoG$ photon subtraction therefore does not lead to a state. The result is stated in the following Lemma.
\begin{lem}
\label{Lemma:PASis0} Let $\rhoG$ be a Gaussian state with covariance matrix $V$ and displacement vector $\boldsymbol{d}$, and let $\boldsymbol{c}\in\C^n$. Then $a(\boldsymbol{c})\rhoG a^\dagger(\boldsymbol{c})=0$ if and only if $\mcbold\in{\mathrm{Ker}}(V-\bbone)$ and $\overline{\boldsymbol m}_{\boldsymbol c}\cdot \boldsymbol{ d}=0$.
\end{lem}
When $V=\bbone$, the Gaussian state is in fact a coherent state $|\zbold\rangle$. In that case the first condition of the Lemma is satisfied for all $\cbold\in\C^n$ and the second condition reads $\overline \cbold\cdot \zbold=0$.  In other words, one has 
\begin{equation}\label{eq:cohzero}
a(\cbold)|\zbold\rangle=0\quad\Leftrightarrow\quad \overline\cbold\cdot \zbold=0.
\end{equation}
Of course, this particular case follows immediately from the well known identity
\begin{equation}\label{eq:cohstateannih}
a(\cbold)|\zbold\rangle=(\overline{\cbold}\cdot \zbold) |\zbold\rangle,
\end{equation}
which is in turn a direct consequence of Eq.~\eqref{eq:pullthrough2}. When there is only one mode, then Eq.~\eqref{eq:cohzero} can only be satisfied if $|\zbold\rangle=|0\rangle$. With several modes, on the other hand it does occur for nonzero $\zbold$.  Lemma \ref{Lemma:PASis0}  treats the case of a general Gaussian state and the proof, which uses Eq.~\eqref{eq:CharacPhotonAddedGaussianState}-\eqref{eq:betapm}, is slightly more involved. \\
\noindent{\bf Proof.} $\tilde \rho^{\mathrm G}_-:=a(\boldsymbol{c})\rhoG a^\dagger(\boldsymbol{c})=0$ if and only if $\tilde \chi^{\mathrm G}_-(\boldsymbol{z})=0$ for all $\zbold\in\C^n$, where $\tilde\chi^{\mathrm G}_-$ is the characteristic function of $\tilde\rho^{\mathrm G}_-$. From Eq.~\eqref{eq:CharacPhotonAddedGaussianState}-\eqref{eq:betapm}, it is given by 
$$
\tilde \chi^{\mathrm G}_-(\zbold)= \left(\frac12
 		\overline{\boldsymbol{m}}_{\boldsymbol{c}}^TV\boldsymbol{m}_{\boldsymbol{c}}\pm\frac12
 		-\boldsymbol{ \beta}^T_\pm\boldsymbol{m}_{\boldsymbol{c}}\overline{\boldsymbol{m}}_{\boldsymbol{c}}^T\boldsymbol{ \beta}_\pm
 		\right)\chiG(\boldsymbol z).
		$$
For this to vanish, the polynomial factor preceding the exponential factor $\chiG$ must vanish for all $\zbold\in\C^n$. Let $\boldsymbol v\in\R^{2n}$ be an eigenvector of $V$ with eigenvalue $\lambda\not=1$. Then define, for all $\mu\in\R$, 
\begin{equation}\label{eq:Zmu}
\boldsymbol{Z}(\mu)=\mu U\Omega^T \boldsymbol v.
\end{equation}
Then 
\begin{equation}\label{eq:betamu}
\boldsymbol \beta_-^T\mcbold=\frac\mu2(\lambda-1)(\Omega \boldsymbol v)^T\mcbold + i (\Omega \boldsymbol d)^T\mcbold.
\end{equation}
It follows that 
\begin{equation}\label{eq:tildechi}
\tilde \chi^{\mathrm G}_-(\boldsymbol{Z}(\mu))=p(\mu)\chiG(\boldsymbol{Z}(\mu)),
\end{equation}
where $p(\mu)$ is a polynomial of degree two. This polynomial vanishes identically if and only if it has vanishing coefficients. One readily checks this is equivalent to  
\begin{eqnarray}
(\Omega \boldsymbol v)^T\mcbold=0,\label{eq:orthogonalmc}\\
\frac12(\overline{\boldsymbol m}_{\boldsymbol c}^T(V-\bbone)\mcbold)+\mid (\Omega d)^T \mcbold\mid^2=0.\label{eq:constant0}
\end{eqnarray}
Since $\Omega^T \mcbold=i \mcbold$, the first of these two conditions is equivalent to $\boldsymbol v^T\mcbold=0$. Since this needs to hold for all eigenvectors of $V$ with eigenvalue $\lambda\not=1$, it follows that $\mcbold\in {\textrm{Ker}}(V-\bbone)$. Hence the first term in Eq.~\eqref{eq:constant0} vanishes and so does therefore the second one. This concludes the proof. \hfill$\square$

We are now ready to fully characterize the classical and hence the nonclassical photon-subtracted Gaussian states. 
\begin{prop}
\label{prop:rhominusclass}Let $\rhoG$ be a Gaussian state. Let $\boldsymbol{c}\in\C^n$ and suppose $a(\cbold)\rhoG a^\dagger(\cbold)\not=0$. Then:
    \setlist{nolistsep}
\begin{enumerate}[label=(\roman*)]
\item \label{prop3}$\rhoGm$ is classical/nonclassical if and only if  $\rhoG$ is classical/nonclassical.
\item\label{prop4} $\rhoGm$ is classical if and only if  $V-\bbone\geq 0$. 
\end{enumerate}
\end{prop}
In Proposition~\ref{prop:rhominusclass}, conditions \ref{prop3} and~\ref{prop4} are equivalent since it is well known that the classicality of a Gaussian state is equivalent to $V\geq \bbone$~\cite{Serafini}. Proposition~\ref{prop:rhominusclass}~(i) asserts that, whereas it is true that photon subtraction cannot produce a nonclassical state from a classical one, it is also true that it does never transform a nonclassical Gaussian state into a classical one. We will show in the next section that it can in fact considerably increase the degree of nonclassicality of a given Gaussian state.

\noindent{\bf Proof.} 
In view of the previous comment, it is sufficient to prove that if $\rhoGm$ is classical then $V\geq \bbone$.  
For that purpose, we use the fact that, if $\rhoGm$ is classical, then the Fourier transform of the $P$-function, which is known to be given by $\e^{\frac12\boldsymbol\xi\cdot\boldsymbol\xi}\chiGm(\boldsymbol\xi)$~\cite{Cahill}, is   a bounded function.  Using Eq.~\eqref{chiGaussian} and~\eqref{eq:CharacPhotonAddedGaussianState} this implies
\begin{eqnarray}
|\e^{\frac12\boldsymbol\xi\cdot\boldsymbol\xi}\chiGm(\boldsymbol\xi)|&=&\Ncal_-{\Big|}\frac12 \overline{\boldsymbol m}_{\boldsymbol{c}}^T V\boldsymbol m_{\boldsymbol{c}}-\frac12-\boldsymbol{\beta}_-^T\boldsymbol m_{\boldsymbol{c}}\overline{\boldsymbol m}_{\boldsymbol{c}}^T\boldsymbol{\beta}_-{\Big|}\nonumber\\
&\ &
\hskip 2cm \times\e^{-\frac12\boldsymbol \xi^T\Omega(V-\bbone)\Omega^T\boldsymbol \xi}\label{eq:explode}
\end{eqnarray}
is bounded. Suppose it is not true that $V\geq \bbone$. Then there exists $\boldsymbol v\in\R^{2n}, \boldsymbol v\cdot \boldsymbol v=1,$ and $0\leq \gamma<1$ so that $V\boldsymbol v=\gamma \boldsymbol v$.  For such $\boldsymbol v$, we define $\boldsymbol Z(\mu)$ as in Eq.~\eqref{eq:Zmu} and hence $\boldsymbol\xi(\mu)=\frac1{\sqrt2}U^\dagger\boldsymbol Z(\mu)=\mu\frac1{\sqrt2}\Omega^T \boldsymbol v$. The exponential factor in~\eqref{eq:explode} then grows without bound for large $\mu$. Hence $\e^{\frac12\boldsymbol\xi\cdot\boldsymbol\xi} \chiGm$ can be bounded only if the polynomial prefactor $p(\mu)$ in Eq.~\eqref{eq:tildechi} vanishes identically.  This in turn is equivalent to Eq.~\eqref{eq:orthogonalmc}-\eqref{eq:constant0}.  Since Eq.~\eqref{eq:orthogonalmc} holds for all eigenvectors of $V$ with eigenvalue strictly less than $1$, it follows that $\mcbold$ belongs to the nonnegative spectral subspace of $V-\bbone$. Eq.~\eqref{eq:constant0} then implies that $\mcbold$ in fact belongs to the kernel of $V-\bbone$. And, in addition, that $\boldsymbol{d}$ is perpendicular to $\mcbold$. By the Lemma, this in turn implies that $a(\cbold)\rhoG a^\dag(\cbold)=0$, which is a contradiction. In conclusion, $V-\bbone\geq 0$. 
\hfill$\square$


\subsection{Wigner positivity/negativity of photon-subtracted Gaussian states}
Using~\eqref{eq:Wignerpm}-\eqref{eq:KpmAc} one easily characterizes the Wigner positive/negative photon-subtracted Gaussian states as follows.
\begin{lem}
	\label{Lemma:WigNegSqThmoins}
Let $\rhoG$ be  a Gaussian state and $\boldsymbol c\in\C^n$. Suppose $\rhoGm$ is Wigner-negative. Then
\begin{equation}\label{eq:WNcrit}
\overline{\mcbold}^TV^{-1}\mcbold >1.
\end{equation}
Suppose that either $\boldsymbol{d}=0$ or  that $1$ is not an eigenvalue of $V$. Then Eq.~\eqref{eq:WNcrit} is both necessary and sufficient for $\rhoGm$ to be Wigner negative. 
\end{lem}
Note that it follows from Lemma \ref{Lemma:WigNegSqThmoins} that
photon-subtracted Gaussian states are Wigner-positive if 
$$
\overline{\mcbold}^TV^{-1}\mcbold\leq 1.
$$
This straightforward condition therefore identifies a family of Wigner-positive states indexed by $V$ and by $\cbold$ which is of independent interest because  a complete characterization of all Wigner positive states is not known~\cite{Brocker}. \\
\noindent{\bf Proof.} Since $\boldsymbol{m}_{\boldsymbol{c}}  \overline{\boldsymbol{m}}_{\boldsymbol{c}}^T$ is a rank one projector in $\C^{2n}$, the term 
$\boldsymbol{ \lambda}^T_- \boldsymbol{m}_{\boldsymbol{c}} \overline{\boldsymbol{m}}_{\boldsymbol{c}}^T\boldsymbol{ \lambda}_-$ in Eq.~\eqref{eq:Wignerpm} is nonnegative. It follows then from Eq.~\eqref{eq:Wignerpm}-\eqref{eq:KpmAc} that if $\rhoGm$ is Wigner negative, then $M_-(V,\boldsymbol c)<0$, which is equivalent to Eq.~\eqref{eq:WNcrit}.  This proves the first statement of  Lemma~\ref{Lemma:WigNegSqThmoins}. For the second statement, note that, if $\boldsymbol{d}=0$, then the term $\boldsymbol{ \lambda}^T_- \boldsymbol{m}_{\boldsymbol{c}} \overline{\boldsymbol{m}}_{\boldsymbol{c}}^T\boldsymbol{ \lambda}_-$ vanishes when $\boldsymbol{r}=0$. When $1$ is not an eigenvalue of $V$, then $(V^{-1}-\bbone)$ is invertible and then this term vanishes provided $\boldsymbol{r}=(\bbone-V)^{-1} \boldsymbol d$.  
Hence, in both these cases, the Wigner function of $\rhoGm$ is negative in at least one point of the phase space  if and only if 
\begin{equation}
M_-(V,\boldsymbol{c})< 0,\label{eq:wignernonneg}
\end{equation} 
which yields the result. \hfill$\square$

In the case where only a single mode is present ($n=1$) the previous result can be sharpened and a link established between the Wigner negativity of the photon-subtracted state and the QCS of the Gaussian mother state. 
First, without loss of generality, one can now take $c=1$ and one finds from Eq.~\eqref{eq:QCSGaussian} that
 \begin{eqnarray}
\Ccal^2(\rhoG)&=&\frac12 \tr V^{-1}=\overline{\boldsymbol{m}}_{\boldsymbol{c}}^T V^{-1} \boldsymbol{m}_{\boldsymbol{c}}, \label{eq:QVSmvm}\\
 M_\pm(V,\boldsymbol{c})&=&\mp\frac12-\frac12\overline{\boldsymbol{m}}_{\boldsymbol{c}}^T V^{-1} \boldsymbol{m}_{\boldsymbol{c}}\nonumber\\
 &=&-\frac12\left(\Ccal^2(\rhoG)\pm 1\right). \label{mcVmcenfctdeQCS} 
 \end{eqnarray} 
Next, introduce an orthogonal eigenbasis $\boldsymbol{e}_1, \boldsymbol{e}_2$ for $V$:
$$
0<v_1\leq v_2, \quad \boldsymbol{ e}_i\in\R^2, \quad V\boldsymbol{ e}_i=v_i\boldsymbol{ e}_i.
$$
One then has the following result.

\begin{lem}
	\label{Lemma:WigNegSqThmoins1mode}
Suppose $v_1>1$. Then the one-mode photon-subtracted Gaussian state $\rhoGm$  is classical and hence Wigner positive. \\
Suppose $v_1<1$. Then the one-mode photon-subtracted Gaussian state $\rhoGm$  is Wigner negative if and only if
\begin{equation}
\Ccal^2(\rhoG)>1.
\end{equation}
Suppose $v_1=1$. Then the one-mode photon-subtracted Gaussian state $\rhoGm$  is Wigner negative if and only if
\begin{equation}
\Ccal^2(\rhoG)>1+(\boldsymbol d^T\boldsymbol{ e}_1)^2.
\end{equation}
In general therefore, if $\rhoGm$ is Wigner negative, then the Gaussian mother state $\rhoG$ is strongly nonclassical, meaning
$\Ccal^2(\rhoG)>1$. 
\end{lem}
We already showed in the previous subsection that photon-subtracted states are nonclassical if and only if $\rhoG$ is nonclassical. One now sees in addition that if their Wigner function has some negativity then the Gaussian mother state is strongly nonclassical. \\
\noindent{\bf Proof.} The first statement follows directly from Proposition~\ref{prop:rhominusclass}. 
Since det$V\geq 1$, the condition $v_1<1$ implies $v_2>1$. Hence Lemma~\ref{Lemma:WigNegSqThmoins} implies the result in this case. 

Now suppose $v_1=1$, so that $v_2\geq 1$. If $v_2=1$, the state $\rhoG$ is a coherent state, in which case $\Ccal(\rhoG)=1$ and the condition is never satisfied; but this is compatible with the statement of the Lemma since, in view of Eq.~\eqref{eq:cohstateannih}, the photon-subtracted state is then the same coherent state and hence Wigner positive. 
 It remains to treat the case where $v_1=1<v_2$. It follows from Eq.~\eqref{eq:Wignerpm} and Eq.~\eqref{eq:lambdapm} that the Wigner function is negative in at least one point if and only if
$$
\min_{\boldsymbol r}\Big(M_-(V,\boldsymbol{c})+	\frac12\|\boldsymbol\lambda_-\|^2
\Big)<0
$$
where
$$
\boldsymbol{ \lambda}_-=\left[\left(V^{-1}-\bbone\right)\boldsymbol{ r}-V^{-1}\boldsymbol{ d}\right]\in\R^{2}.
$$
Since
$
\min_{\boldsymbol r}\|\boldsymbol\lambda_-\|^2=(\boldsymbol{e}_1^T\boldsymbol d)^2
$
the result then follows from Eq.~\eqref{mcVmcenfctdeQCS}.
\hfill$\square$

In the next sections we turn to a quantitative analysis of the Wigner negativity and the QCS for single-mode photon-added/subtracted squeezed thermal states. We will show that the Wigner negativity of such states is bounded above by that of the one-photon Fock state, and very sensitive to noise and squeezing. The QCS can on the other hand be very strongly enhanced by the photon-addition/subtraction process and increases with the squeezing. It is also sensitive to noise and losses.



\section{Photon-added squeezed thermal states}\label{s:SqThPlus}
In this section, we quantitatively evaluate the effect produced by adding a photon to a general centered single-mode Gaussian state on the Wigner negative volume and on the QCS of the state.

We note that the nonclassical nature of  such photon-added states has previously been certified theoretically and/or experimentally only for the two particular cases of photon-added thermal states (see \cite{Agarwal2,Devi,Kiesel}) and of photon-added squeezed vacuum states (see \cite{Fabre,FabrePRL17, hu,Kun}) using various nonclassicality witnesses, but without providing a complete quantitative assessment, even in these particular cases. 

Our analysis of the Wigner negative volume of the photon-added Gaussian states shows that it is highest for photon-added squeezed vacuum states. It is sensitive to noise and, in the presence of noise, it decreases with increased squeezing (Sec.~\ref{s:WNSqTh+}). In this sense, there is -- at a fixed noise level -- a tradeoff between Wigner negativity and squeezing for such states. We will further see that the degaussification process of photon-adding tends to entail a considerable percentage increase in QCS (Sec.~\ref{s:QCSSqTh+}). Whereas this entails a corresponding gain in nonclassicality, it also means the resulting state is considerably more sensitive to environmental decoherence than its Gaussian mother state, as explained in Section~\ref{s:QCS}. 

A squeezed thermal (SqTh) state is defined as 
\begin{equation}
\rho_{\text{SqTh}}=S \rho_{\text{Th}} S^\dag \ \textrm{where}\  
\rho_{\text{Th}}=(1-q)\sum_n q^n\ket{n}\bra{n}\label{eq:SqTh}
\end{equation}
is a thermal state of temperature\footnote{The actual temperature is given by $T$ with $q=e^{\frac{-\hbar\omega}{kT}}$; $q$ is also related to the mean photon number $\mean{n}$ as $q=\frac{\mean{n}}{1+\mean{n}}$. } $q$ and $S=~e^{\frac12(\bar z a^2-z a^{\dag 2})}$ is the squeezing operator with $z=re^{i\phi}$. The rotational invariance of the QCS implies we can restrict ourselves to the case where $\phi=0$. 
\begin{figure*}[!t]
	\centering
	 \begin{subfigure}[b]{0.45\textwidth}
	\centering
	\includegraphics[height=7.18cm,keepaspectratio]{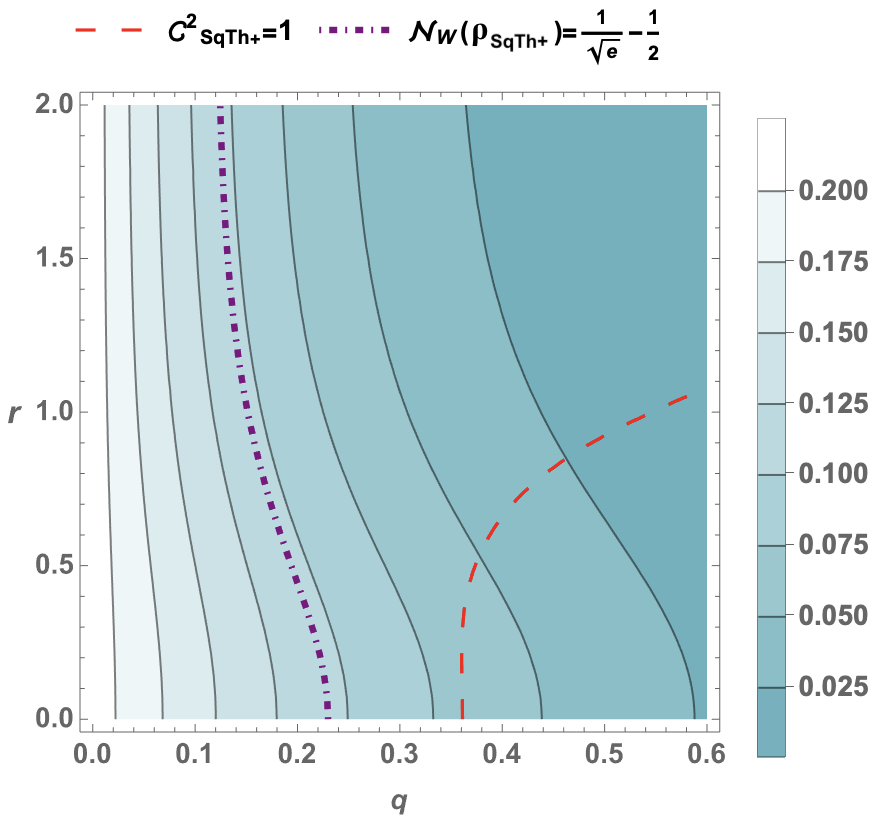}
	\caption{
	}
	\label{fig:NegWSqThPlus}
\end{subfigure}
	 \begin{subfigure}[b]{0.42\textwidth}
		\centering
		\includegraphics[height=7.2cm,keepaspectratio]{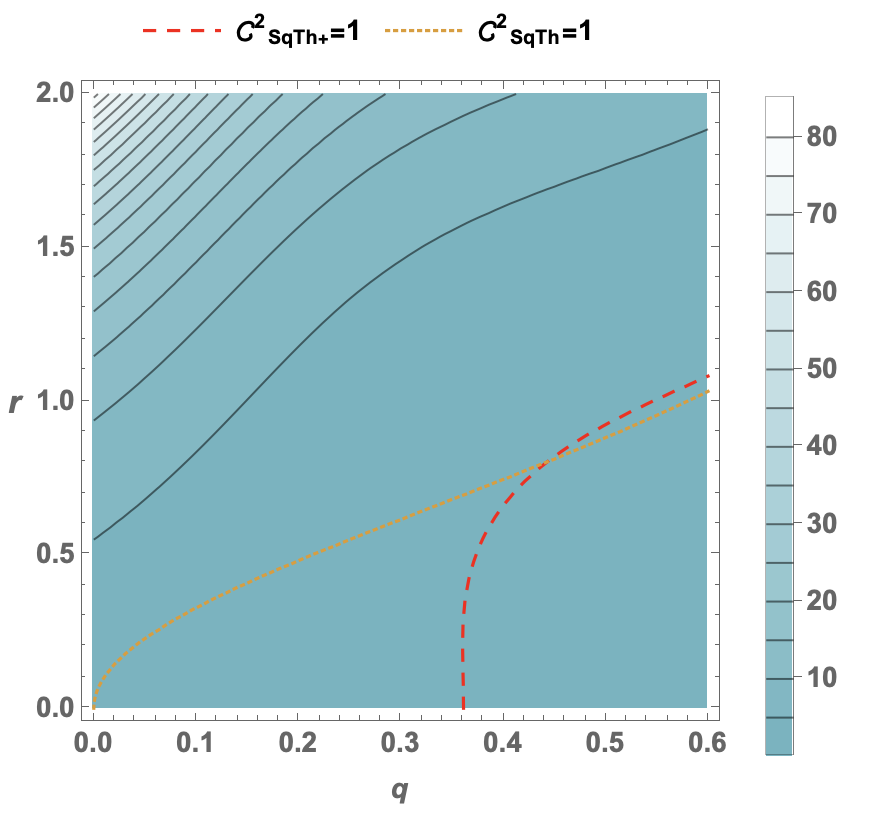}
		\caption{
		}
		\label{fig:ContourPlotSqTh+}
	\end{subfigure}
	 \begin{subfigure}[b]{0.42\textwidth}
	\centering
	\includegraphics[height=7.1cm,keepaspectratio]{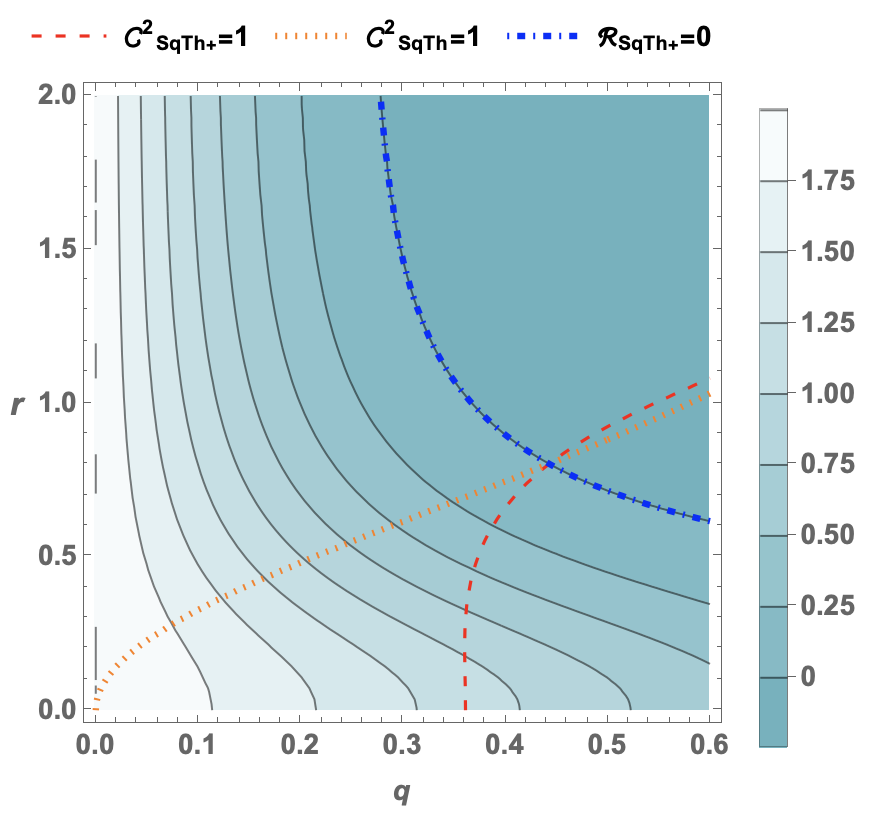}
	\caption{
	}
	\label{fig:ContourPlotR+}
\end{subfigure}
	\caption{Level lines of (a) the Wigner negative volume $N_{\text{W}}(\rho_{\text{SqTh}+})$, (b) the QCS, $\mathcal{C}^2_{\text{SqTh}+}$ and (c) the relative gain  $\mathcal{R}_{\text{SqTh}+}$  of photon-added squeezed thermal states in function of the temperature $q$ and the squeezing $r$.  In dashed red the line $\mathcal{C}^2_{\text{SqTh}+}(q,r)=1$. In dotted orange, the level line $\mathcal{C}^2_{\text{SqTh}}(q,r)=1$ of the QCS of the squeezed thermal states. \label{fig:SqThp}}
\end{figure*}
The covariance matrix of these states is
\eqstar{V_{\text{SqTh}}=\frac{1+q}{1-q}\begin{psmallmatrix}e^{-2r}&0\\0&e^{2r}
\end{psmallmatrix}}
and their characteristic function is
\eqstar{\chi_{\text{SqTh}}(\boldsymbol z)=e^{-\frac12\frac{1+q}{1-q} \left(e^{2 r}  \xi_1^2+e^{-2 r}  \xi_2^2\right)},}
where we recall $\boldsymbol z=\xi_1+i\xi_2$. Their QCS, computed with Eq. \eqref{eq:QCSGaussian}, is then equal to
 \eq{\Ccal^2_{\text{SqTh}}(q,r)=\frac{1-q}{1+q}\cosh(2r).\label{eq:QCSSqTh}}
 Note that it increases sharply with the squeezing parameter $r$ and decreases with $q$. Increased squeezing therefore reduces the decoherence time sharply. 
A photon-added squeezed thermal ($\text{SqTh}+$) state is defined as 
\eqstar{\rho_{\text{SqTh}+}=\mathcal{N}_{\text{SqTh}+}\,a^\dag \rho_{\text{SqTh}} a}
where $\mathcal{N}_{\text{SqTh}+}=2\left(1+\frac{1+q}{1-q}\cosh(2r)\right)^{-1}$. Its characteristic function can be computed using Eq.~\eqref{eq:CharacPhotonAddedGaussianState}:
\eqarray{\chi_{\text{SqTh}+}(\boldsymbol z)&=&\chi_{\text{SqTh}}(\boldsymbol z)\left(\frac{2 q |\boldsymbol  z|^2}{\left(1-q^2\right) \cosh 2 r+(1-q)^2}\right.\nonumber\\
	&&\left.+\frac{q+1}{q-1} \left(e^{2 r}  \xi_1^2+e^{-2 r}  \xi_2^2\right)+1\right).\label{ChiSqThPlus}}
\subsection{The Wigner negativity of photon-added squeezed thermal states}\label{s:WNSqTh+}
To evaluate  the Wigner negativity of the $\text{SqTh}+$ states, we evaluate, as is customary, their  Wigner negative volume~\cite{Kenfack}, that we shall denote by $N_{\text{W}}(\rho)$: it is defined as  the absolute value of the integral of the Wigner function over the area where the latter is negative. We recall that the Wigner negative volume has been proven to be  a (non-faithful) monotone of genuine (or quantum) non-Gaussianity~\cite{Takagi18}. The Wigner function of the $\text{SqTh}+$ state can be readily computed with \eqref{eq:Wignerpm} (see \ref{Appendix E} for the details). One sees that 
it is negative inside an ellipse centered at the origin where it reaches its minimal value. Except when $q=0$, a general analytical expression for $N_{\text{W},\text{SqTh}+}(q,r)$ is not readily obtained, but the result of a numerical computation  is shown in Fig.~\ref{fig:NegWSqThPlus}. 

For SqV+ states, when $q=0$, an analytical computation yields the following value, independently of $r$~\cite{Kenfack}:
\begin{equation}\label{eq:NWSqVp}
N_{\text{W},\text{SqV}}(r)=N_{\text{W},\text{SqTh}+}(0,r)=\frac{2}{\sqrt{e}}-1 = 0.213.
\end{equation}
This is the maximal value attained on $\text{SqTh}+$ states, and it is in particular the value for the first Fock state.

The Wigner negative volume $N_{\text{W},\text{SqTh}+}(q,r)$ 
decreases with the noise $q$, at a given value of the squeezing $r$: this is not surprising, since higher $q$ is expected to make the state more classical. The dot-dashed purple line on Fig. \ref{fig:NegWSqThPlus} indicates for which value of the noise the Wigner negative volume drops down to half the value it takes on the first Fock state. This happens with a noise in the range  $0.12\leq q \leq 0.2$ depending on the squeezing and it shows  that the Wigner negative volume of the $\text{SqTh}+$ states is quite sensitive to noise.

Contrary to what happens when $q=0$, when $q\not=0$,  the Wigner negative volume does depend on the squeezing and in fact decreases with increasing $r$. In this sense, at a given noise level, there is a tradeoff to be considered: one pays in Wigner negative volume to gain in squeezing.  In addition, as we will see below, increased squeezing substantially reduces the decoherence time. 

The Wigner negative volume saturates to a finite value at large $r$ that decreases with $q$ and 
that is readily computed (see \ref{Appendix E}) for small $q$:
\begin{eqnarray}
N_{\text{W},\text{SqV}}(r)=N_{\text{W},\text{SqV}}(+\infty)&\geq &N_{\text{W},\text{SqTh}+}(q,+\infty)\nonumber\\
&\approxeq&\left(\frac{2}{\sqrt{e}}-1\right)\frac{(1-q)^3}{(1+q)^3}.
\nonumber
\end{eqnarray}

	\subsection{The QCS of photon-added squeezed thermal states}\label{s:QCSSqTh+}
The QCS of the $\text{SqTh}+$ states can be computed with Eq.~\eqref{eq:QCSWch}.
 The result is explicit (it can be found in  \ref{Appendix D bis}), but is not very instructive. To second order in $q,r$, it reads
 \begin{equation}
 \Ccal_{\text{SqTh}+}^2(q,r)\approxeq3-8 q+8 q^2+6 r^2\nonumber
 \end{equation}
 The expression simplifies considerably for photon-added squeezed vacuum (SqV+, $q=0$) states and for photon-added thermal (Th+, $r=0$) states: 
 \eqarray{\Ccal_{\text{SqV}+}^2(0,r)&=&3\cosh(2r)\label{eq:QCSSqVp}\\
	\Ccal_{\text{Th}+}^2(q,0)&=&\frac{6}{q+1}-1-\frac{2 (q+1)}{q^2+1}.\label{eq:QCSThp}}

Examining Fig.~\ref{fig:ContourPlotSqTh+}, one observes that the QCS  of $\text{SqTh}+$ states increases sharply with $r$ and decreases with $q$, as the QCS of their Gaussian mother states (see Eq.~\eqref{eq:QCSSqTh}). 
The QCS of the $\text{SqTh}+$ states tends however to be considerably higher as we will see below. While they therefore display correspondingly stronger nonclassical effects, they are also more prone to environmental decoherence.  For example, the QCS (squared) of the first Fock state (which is the photon-added state of the vacuum, corresponding to $q=0=r$) equals $3$ [see Eq.~\eqref{eq:FockQCS}], while that of the vacuum itself is only $1$: this corresponds to a 200$\%$ increase.

We now investigate quantitatively how strongly the degaussification through photon-addition affects the QCS for general $(q,r)$.   
For that purpose we will use the relative QCS change $\mathcal{R}_\pm(\rho)$ defined as
		\eq{\mathcal{R}_\pm(\rho)=\frac{\Ccal^2({\rho_\pm})-\Ccal^2({\rho})}{\Ccal^2({\rho})},\label{RelativeGain}}
so that $\Ccal^2(\rho_\pm)=(1+\Rcal_\pm(\rho))\Ccal^2(\rho)$.  It provides the percentage change in QCS as a result of the photon-addition/subtraction process. It is indeed clear that some of the QCS of the photon-added/subtracted Gaussian states is inherited from the Gaussian mother state to which a photon is added or from which it is subtracted; and that part of it is due to the addition/subtraction process itself. 

We show in Fig. \ref{fig:ContourPlotR+} the contour plot  in the $(q,r)$-plane of the relative change $\mathcal{R}_{\text{SqTh}+}(q,r)$ of the QCS obtained with the addition of a photon. From Eq.~\eqref{eq:QCSGaussian} and Eq.~\eqref{eq:QCSSqVp} one sees that  for squeezed vacuum states, one has $\mathcal{R}_{\text{SqTh}+}(0,r)=2$, independently of the squeezing. This corresponds to a $200\%$ increase of the QCS due to photon addition, and it is the maximal value reached, as can be seen from the figure. When $q>0$, the change in QCS  decreases with increasing $r$ and with increasing $q$. Nevertheless, there is a large region in the parameter space $(q,r)$ where the relative change is positive and sizable.   For $q<0.1$ and values of $r$ in the range $1\leq r\leq 2$ (which corresponds to a squeezing factor comprised between 7 and 15 dB), it is at least 90\%, for example.  For $q<0.2$ and the same range of $r$ values, it is still at least $50\%$.

In the region to the right of the blue dot-dashed curve, the relative gain is negative. This means that photon-addition leads to a decrease in QCS and a concomitant increase in decoherence time. The latter is however less than $10\%$ in the region represented. In addition, in this region the Wigner negative volume of the states is small, at most 25$\%$ of the maximal value reached on the photon-added squeezed vacuum states. 

	\begin{figure*}[!t]
	\centering
		\begin{subfigure}[b]{0.45\textwidth}
		\centering
		\includegraphics[height=7cm,keepaspectratio]{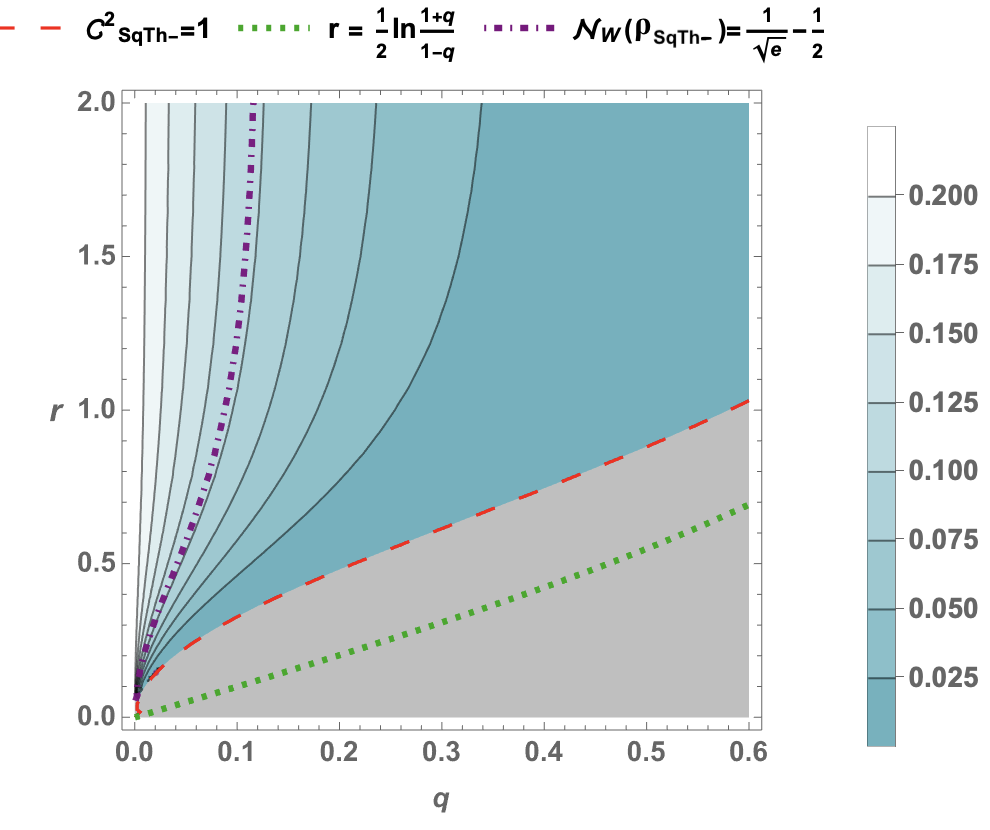}
		\caption{
		}
		\label{fig:NegWSqThMoins}
	\end{subfigure}
	\begin{subfigure}[b]{0.42\textwidth}
		\centering
		\includegraphics[height=7.1cm,keepaspectratio]{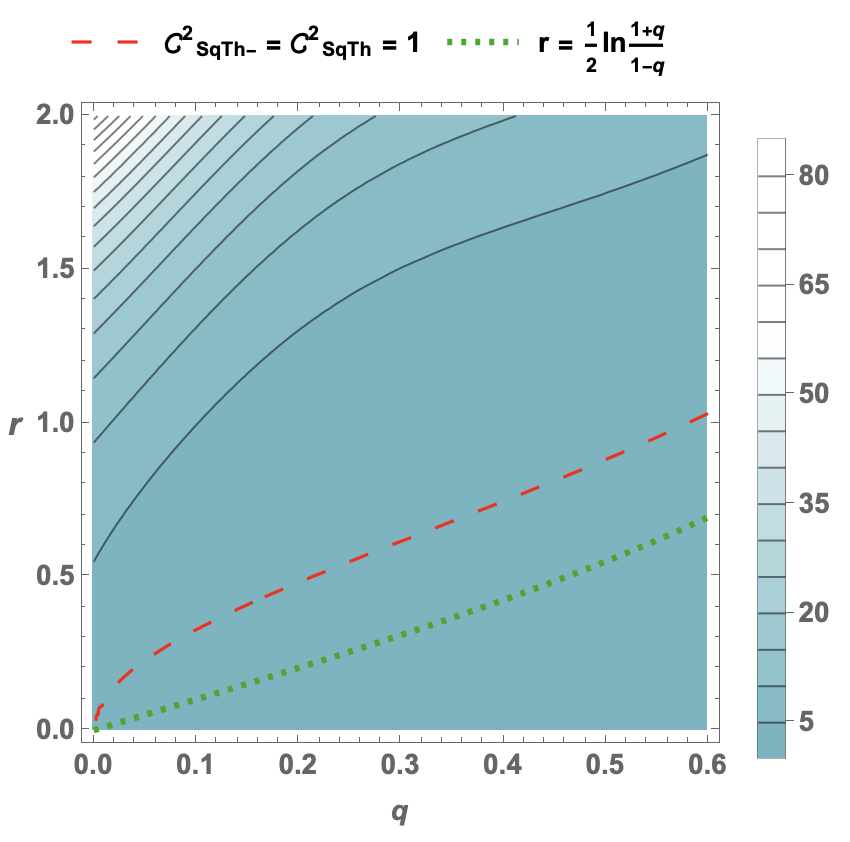}
		\caption{
		}
		\label{fig:ContourPlotSqTh-}
	\end{subfigure}
	\begin{subfigure}[b]{0.42\textwidth}
		\centering
		\includegraphics[height=7.1cm,keepaspectratio]{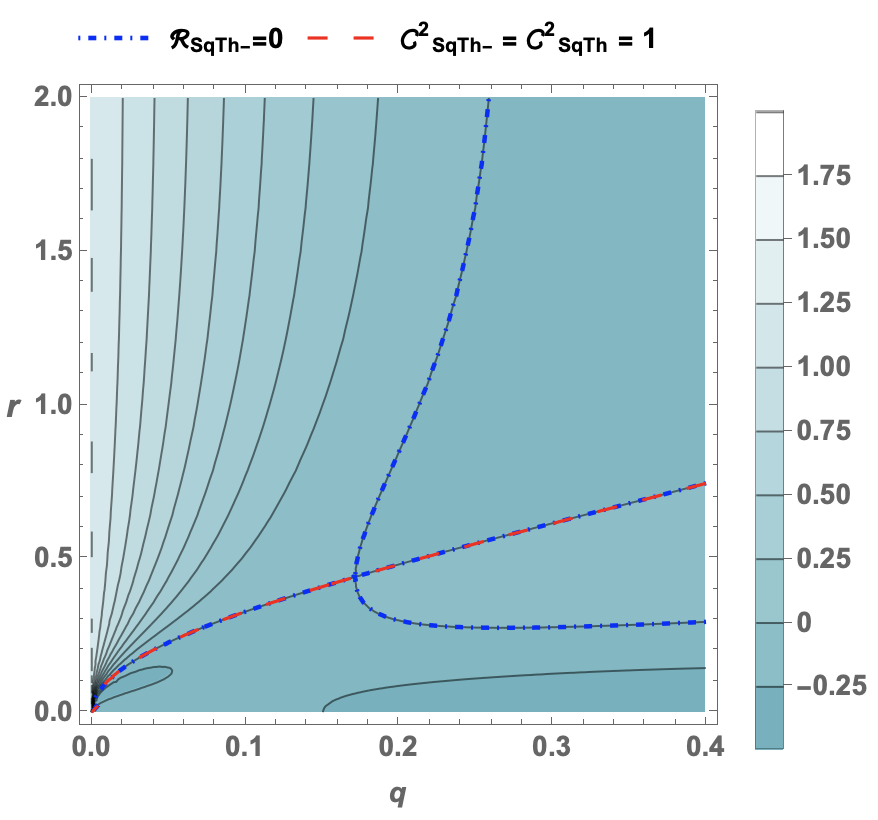}
		\caption{
		}
		\label{fig:ContourPlotSqThMoins}
	\end{subfigure}
	\begin{subfigure}[b]{0.42\textwidth}
	\centering
	\includegraphics[height=7.1cm,keepaspectratio]{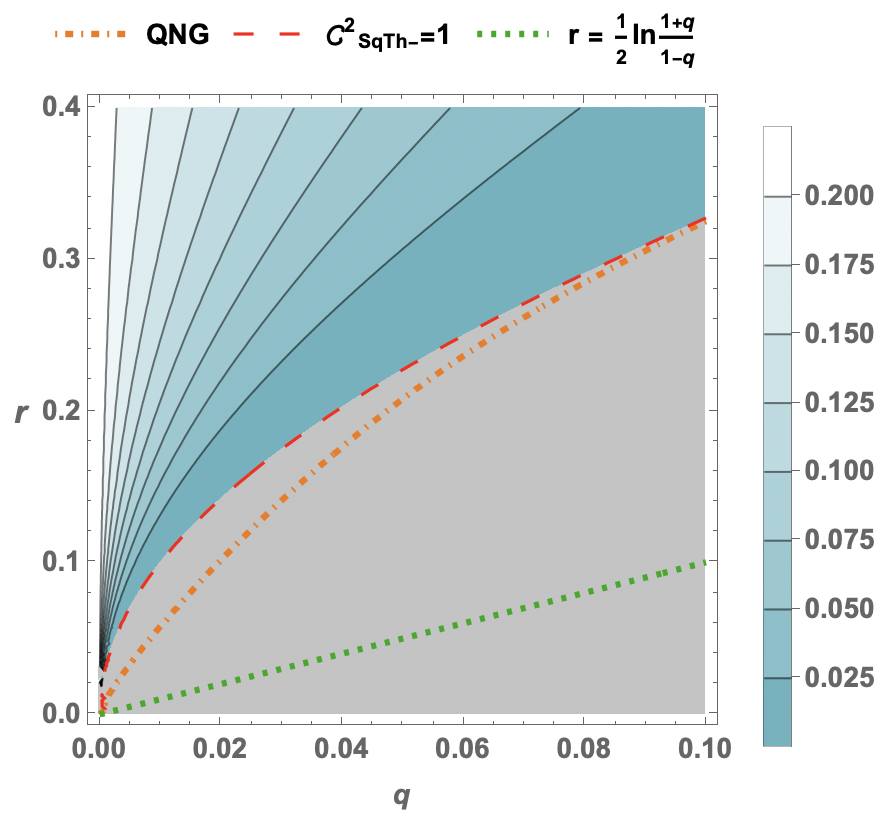}
	\caption{
	}
	\label{fig:QNG_for_SqThMoins}
\end{subfigure}
	\caption{  Level lines of (a) the Wigner negative volume $N_{\text{W}}(\rho_{\text{SqTh}-})$, (b) the QCS $\mathcal{C}^2_{\text{SqTh}-}$, and (c) the relative gain $\mathcal{R}_{\text{SqTh}-}$  of photon-subtracted squeezed thermal states in function of the temperature $q$ and the squeezing $r$.   Panel (d) shows a zoom of (a) where the dashed-dotted orange line indicates the  values of $q$ and $r$ where the inequality Eq.~\eqref{eq:qng} is saturated. States above this curve are quantum non-Gaussian.	In dashed red the line $\mathcal{C}^2_{\text{SqTh}-}(q,r)=\mathcal{C}^2_{\text{SqTh}}(q,r)=1$, and in dotted green,  the line $r=\frac12\ln(\frac{1+q}{1-q})$  below which the SqTh and $\text{SqTh}-$ states are classical. 
Above the dashed red line, both types of states are strongly nonclassical and the $\text{SqTh}-$ states have Wigner negativity.  In the gray region, the Wigner function is positive. 	
	The region delimited by the dotted green and dashed red lines corresponds to weakly nonclassical states.  
\label{fig:SqThm}}
\end{figure*}

Finally, one may note that the level curves of $\mathcal{R}_{\text{SqTh}+}(q,r)$ have vertical asymptotes, reflecting the fact that, at large $r$, the change in QCS is independent of the squeezing. One finds readily, for all $q$ and $r$ (see \ref{Appendix D bis}) that 
\eq{ \mathcal{R}_{\text{SqTh}+}(q,r)\geq\mathcal{R}_{\text{SqTh}+}(q, +\infty)=2-12q\frac{q^2+1}{q^4+10q^2+1}.
\nonumber}
 For example, when $q\leq 0.1$, it is larger than $90\%$ for all values of $r$. In view of what precedes, one observes that this asymptotic value is nearly reached when $r=2$.
 We conclude that a considerable increase of the QCS can therefore result from the photon-addition process at experimentally accessible values of the squeezing and provided $q$ is not too large. 

In conclusion, if the Wigner negative volume is used as the figure of merit, degaussification of a Gaussian one-mode state through photon addition gives an optimal result for squeezed vacua, independently of the amount of squeezing. This means that photon addition can both produce a Wigner negativity equal to that of a one-photon Fock space and at the same time admit an arbitrary amount of squeezing. However, as our analysis shows, the higher the squeezing, the more sensitive the Wigner negative volume is to noise, which is always present.  In addition, a high squeezing induces a large QCS in the $\text{SqTh}+$ states, meaning that the resulting states are very sensitive to environmental decoherence, much more so than their Gaussian mother states.



\section{Photon-subtracted squeezed thermal states}\label{s:SqThminus}
Like in the photon-addition case, subtracting a photon can enhance certain nonclassical features of the state, as already mentioned in \cite{Fabre, FabrePRL17, Kim}. 
We provide here a quantitative analysis of  the Wigner negative volume and QCS of photon-subtracted squeezed thermal ($\text{SqTh}-$) states  and compare the results with those of the previous section. Details of the computations, which follow along similar lines as those for the $\text{SqTh}+$ states, can be found in~\ref{Appendix D bis} and~\ref{Appendix E}. 

As already mentioned, photon-subtraction turns the one-photon Fock state into the vacuum whereas photon-addition turns it into the two-photon Fock state. Photon-subtraction also preserves the coherent states and generally transforms a classical state into a classical state. This suggests that photon subtraction reduces the nonclassical nature of any state to which it is applied or at least that it cannot be very efficient in enhancing it. Whereas photon-addition increases the nonclassicality efficiently. By investigating the $\text{SqTh}-$ states we will see this is  indeed correct, but only to some extent. We will distinguish three  regimes: $q=0$, $q\not=0$ and $r$ small, $q\not=0$  and $r$ large.

In the absence of noise ($q=0$), it is well known that adding a photon to or removing a photon from a squeezed vacuum (SqV) state (with $r>0$) produces in fact the exact same state. Indeed, using the relations
\eqarray{S^\dag(z)a^\dag S(z)&=&a^\dag\cosh r-ae^{-i\phi}\sinh r,\nonumber\\
S^\dag(z)a S(z)&=&a\cosh r-a^\dag e^{i\phi}\sinh r,\nonumber}
we have
\eqarray{a^\dag\ket{\text{SqV}}&=&a^\dag S\ket{0}=S(a^\dag\cosh r-ae^{-i\phi}\sinh r)\ket{0}\propto S\ket{1},\nonumber\\
a\ket{\text{SqV}}&=&a S\ket{0}=S(a\cosh r-a^\dag e^{i\phi}\sinh r)\ket{0}\propto S\ket{1}.\nonumber}
Hence, once they are normalized, the SqV+ and $\text{SqV}-$ states are identical. They therefore have the same Wigner negative volume [see Eq.~\eqref{eq:NWSqVp}] and the same QCS [see Eq.~\eqref{eq:QCSSqVp}], both independent of $r$. In the absence of noise, photon-subtraction is consequently not less efficient than photon addition in creating nonclassical features. 

We now consider the case where $q\not=0$. In that case, the photon-added and -subtracted states are distinct. We plot the Wigner negative volume of the $\text{SqTh}-$ states in Fig.~\ref{fig:NegWSqThMoins} and their QCS in Fig.~\ref{fig:ContourPlotSqTh-}.   Recall first from Proposition~\ref{prop:rhominusclass} that the line  (dotted  green) 
$$
r=\frac12\ln\left(\frac{1+q}{1-q}\right)
$$
separates the classical SqTh states from the nonclassical ones and also the classical $\text{SqTh}-$ states from the nonclassical ones. So, $\text{SqTh}-$ states are nonclassical only if sufficiently squeezed. This is in contrast with $\text{SqTh}+$ states, that always are Wigner negative and hence nonclassical. In addition, for the $\text{SqTh}-$ state to be Wigner negative, the squeezing must be larger still: the point $(q,r)$ must lie above the curve $\Ccal^2_{\text{SqTh}-}(q,r)=1$ (red dashed), which   can be proven to coincide with the curve $\Ccal^2_{\text{SqTh}}(q,r)=1$. In the region between the (red) dashed and (green) dotted curves, one therefore finds nonclassical Wigner positive states. They are weakly nonclassical since $\Ccal^2_{\text{SqTh}-}(q,r)<1$; note that photon-subtraction therefore transforms a weakly nonclassical Gaussian state into a weakly nonclassical photon-subtracted state. We conclude that, in the presence of noise and at low enough squeezing, the $\text{SqTh}-$ states are either classical or else weakly nonclassical and Wigner positive. More generally, in comparing photon-addition to photon-subtraction, we find that, for all values of $q$ and $r$,
 $$
 N_{\text{W},\text{SqTh}-}(q,r)\leq N_{\text{W},\text{SqTh}+}(q,r).
 $$
We now turn to the question of the (quantum) non-Gaussianity of the photon-subtracted Gaussian states. It is guaranteed to hold whenever the squeezing is strong enough so that the state is strongly nonclassical, since then the Wigner volume of those states does not vanish. When the squeezing is too small, they are classical and hence in the convex hull of the Gaussian states.    The question therefore poses itself nontrivially only for the weakly non-classical photon-subtracted Gaussian states that correspond to the points in the region between  the (red) dashed and (green) dotted curves in Fig.~\ref{fig:SqThm}, which are Wigner positive. We will use the sufficient criterium for quantum non-Gaussianity developed in~\cite{Genoni13} to address the question. It is shown in~\cite{Genoni13} that, if a state's Wigner function satisfies
\begin{equation}\label{eq:qng}
W(0)\leq \frac{2}{\pi} \e^{-2\overline n(1+\overline n)}
\end{equation}
then the state is quantum non-Gaussian. Here $\overline n=\Tr(\rho a^\dagger a)$ is the mean photon number of $\rho$. For the $\textrm{SqTh}-$ states under consideration here, we have (see \ref{Appendix D})
$$
\WGm(0)=\mathcal N_- M_-(V)\WG(0)=\frac1{\bar{n}^{\textrm G}} M_-(V) \frac2{\pi}\frac1{\sqrt{\det V}},
$$
where
$
\mathcal N_-=\frac1{\Tr \rho^{\textrm G} a^\dagger a}=\frac1{\bar{n}^{\textrm G}}
$
and $\bar{n}^{\textrm G}$ is  the mean photon number of the Gaussian mother state and where
$$
M_-(V)=\frac12(1-\mathcal C^2(\rho^{\textrm G})).
$$
This yields explicitly
$$
\WG(0)=\frac{2(1-q)^2(1+q-(1-q)\cosh(2r))}{\pi (1+q)^2((1+q)\cosh(2r)-(1-q))} 
$$
and
\eqarray{
\bar{n}_-&=&\tr \rho_{\text{SqTh}-}a^\dag a\\
&=&\int \WGm(\boldsymbol{\alpha})\left(\alpha_1^2+\alpha_2^2-\frac12\right)d^2\boldsymbol{\alpha}\nonumber\\
&=&\frac12\left(\frac{3(1+q)\cosh(2r)-\frac{4q}{(1+q)\cosh(2r)-(1-q)}}{1-q}-1\right).\nonumber
}
The points where the inequality in Eq.~\eqref{eq:qng} are saturated are indicated as the (orange) dashed-dotted line in Fig.~\ref{fig:SqThm}~(d). Above this line, and below the (red) dashed line,  the states are therefore guaranteed to be  quantum non-Gaussian. Finer criteria would be needed to decide if the states between the (orange) dashed-dotted line and above the (green) dotted line are quantum non-Gaussian.

We may conclude that at low squeezing, photon-addition applied to Gaussian states creates Wigner negativity whereas  photon-subtraction does not and, in general, that the Wigner negative volume is larger after photon-addition than after photon-subtraction. This indicates that photon-addition is more efficient in inducing nonclassical features, a picture that is confirmed by the analysis of the QCS at small and intermediate values of $r$ that follows. As we shall see however, the relative advantage of photon-addition over photon-subtraction is strongly suppressed at large squeezing. 

As in the case of photon-addition, the explicit expression for the QCS is not very instructive for general $q$ and $r$ (see \ref{Appendix D bis}), but it simplifies for SqV$-$ and $\text{SqTh}-$ states to
\begin{eqnarray}
\Ccal_{\text{SqV}-}^2(0,r)&=&3\cosh(2r)=\Ccal^2_{\textrm{SqV}+}(0,r),\nonumber\\
\Ccal_{\text{Th}-}^2(q,0)&=&\frac{6}{q+1}-3-\frac{2 (1-q)}{q^2+1}\leq 1.\nonumber
\end{eqnarray}
 The QCS of $\text{SqTh}-$ states  is plotted in Fig.~\ref{fig:ContourPlotSqTh-}. One sees that, as for $\text{SqTh}+$ states,  $\mathcal{C}^2_{\text{SqTh}-}(q,r)$ is  increasing in $r$ and  decreasing in $q$. 
 
 Comparing the effect of  photon-addition on the QCS  to the one of photon-subtraction,  we find that, provided either $q<0.5$ or $r<0.5$, 
\begin{equation*}\label{eq:compareQCS}
\Ccal_{\text{Th}-}^2(q,r)\leq \Ccal_{\text{Th}+}^2(q,r).
\end{equation*}
 The last inequality is reversed  when $q>0.5$ and $r>0.5$ but in this region the nonclassical features of the photon-added/subtracted states are at any rate limited as can be seen from Fig.~\ref{fig:SqThp}-\ref{fig:SqThm}.  This again indicates that photon-addition tends to enhance the nonclassical features more than photon-subtraction. For example, one finds $\Ccal_{\text{SqTh}+}^2(0.1, 0.5)\approxeq 3.12$
and
$$
\Ccal_{\text{SqTh}-}^2(0.1, 0.5)\approxeq 0.5,\ \Ccal_{\text{SqTh}+}^2(0.1, 0.5)\approx 1.55.
$$
Similarly $N_{\text{W},\text{SqTh}+}(0.1, 0.5)\approxeq 0.15$, and
$$
 N_{\text{W},\text{SqTh}-}(0.1, 0.5)\approxeq 0.23,\ N_{\text{W},\text{SqTh}+}(0.1, 0.5)\approxeq 0.034.
$$
 Note that, at these values of $q$ and $r$, the Wigner negative volume of the $\text{SqTh}-$ state represents only 16$\%$ of the maximal possible value, which is the one of the SqV$\pm$ state ($N_{\text{W},\text{SqV}\pm}=0.213$). The Wigner negative volume of the $\text{SqTh}+$ state is still 70$\%$ of the maximal value.  This is due to a general effect, namely that the loss of Wigner negativity  due to the noise, at a given value of $r$, is  larger for $\text{SqTh}-$ states: the level lines of the Wigner negative volume  are closer together (compare Fig.~\ref{fig:NegWSqThPlus} and Fig.~\ref{fig:NegWSqThMoins}). As a result, for a given squeezing parameter $r$,  the Wigner negative volume of a $\text{SqTh}-$ state drops down to half its value for the single photon state (purple dot-dashed line) at a smaller $q$ value than for the corresponding $\text{SqTh}+$ state.    
 
 We conclude that, whereas at $q=0$, photon-addition and -subtraction applied to Gaussian states produce exactly the same result, the nonclassical properties of those states -- and in particular their Wigner negative volume -- are more sensitive to noise for the case of photon-subtraction than for the one of photon-addition. On the other hand, the price to pay is that the photon-added states, having a larger QCS, are more sensitive to decoherence. 
 
We finally consider the regime where $r$ is very large. The situation is then very different. For the Wigner negative volume, one finds
\begin{eqnarray}
N_{\text{W},\text{SqTh}+}(q,+\infty)&=&N_{\text{W},\text{SqTh}-}(q,+\infty)\nonumber\\
&\approxeq&\left(\frac{2}{\sqrt{\textrm{e}}}-1\right)\frac{(1-q)^3}{(1+q)^3};
\nonumber
\end{eqnarray}
it is identical for photon-added and for photon-subtracted Gaussian states. 

In addition,
we plotted in Fig. \ref{fig:ContourPlotSqThMoins} the relative gain $\mathcal{R}_{\text{SqTh}-}$ of the photon subtracted squeezed thermal state. 
As for photon addition, the level curves of the relative gain have vertical asymptotes meaning that at large $r$ the gain is independent of the squeezing. This asymptotic value is again identical for photon-addition and for photon-subtraction and for the latter it now upper bounds the relative gain;
\begin{eqnarray}
\mathcal{R}_{\text{SqTh}-}(q,r)\leq\mathcal{R}_{\text{SqTh}\pm}(q,+\infty)&=&2-\frac{12 q \left(q^2+1\right)}{q^4+10 q^2+1}\nonumber\\
&\leq & \mathcal{R}_{\text{SqTh}+}(q,r).
\nonumber
\end{eqnarray}
This means that at large enough $r$ a sizable relative gain in QCS  is observed when the state is not too noisy, both for photon-addition and for photon-subtraction.


In fact, by noticing that $\mathcal{R}_{\text{SqTh}+}(+\infty, q)=\mathcal{R}_{\text{SqTh}-}(+\infty, q)$, we see that photon addition or subtraction has a very similar effect on sufficiently squeezed Gaussian states. In this regime, we therefore find the following simple approximate formula for the QCS of either states:
$$
\Ccal^2_{{\textrm{SqTh}}\pm}(r,q)\simeq \left(3-\frac{12 q \left(q^2+1\right)}{q^4+10 q^2+1}\right)\frac{1-q}{1+q}\cosh(2r).
$$
The error between this formula and the exact expression is less than 10$\%$ for all values of $q$ and for $r>1.3$ in the photon-added case and for $r>0.6$ in the photon-subtracted case.

In fact, as shown in~\ref{Appendix E}, in the limit $r\to+\infty$, the photon-added and photon-subtracted Gaussian states coincide at all values of $q$.



\section{Photon-added two-mode Gaussian states}\label{s:2mode}
In this section we illustrate how the general formulas for the Wigner and characteristic functions of photon-added Gaussian states shown in Sec.~\ref{gaussiancase} can be used to study their nonclassical features in the case when two modes are present. We will consider states of the form
\begin{equation}\label{eq:twomodep}
a^\dagger(\boldsymbol{c})\rhoG\otimes\rhoG a(\boldsymbol{c}),
\end{equation}
where $\rhoG$ is a single-mode Gaussian state. We will not give an exhaustive treatment here, but consider two particular cases. In Sec.~\ref{s:2cohp} we consider the case where $\rhoG$ is a coherent state. Such states where realized experimentally as reported in~\cite{Biagi2020}.  In Sec.~\ref{s:2modesqth+} we consider the case where $\rhoG$ is a squeezed thermal state. 
\subsection{Photon added two-mode coherent state }\label{s:2cohp}
In \cite{Biagi2020}   the delocalized single-photon addition on two input modes containing identical coherent states $\ket{\alpha}$  is realized experimentally.
Two families of states are thus constructed and studied:
\eq{\ket{\psi_{\overset{\textrm{even}}{\textrm{\tiny odd}}}}=\frac{\mathcal{N}_{\overset{\textrm{even}}{\textrm{\tiny odd}}}}{\sqrt{2}}(a_1^\dag\ket{\alpha}\ket{\alpha}\pm\ket{\alpha}a_2^\dag\ket{\alpha})\nonumber}
where   $\mathcal{N}_\textrm{even}=(1+2|\alpha|^2)^{-1/2}$ and $\mathcal{N}_\textrm{odd}=1$. 

The Wigner negative volume of these states can be computed from Eq.~\eqref{eq:Wignerpm}. One finds
\eqarray{
		N_{\text{W}_\textrm{even}}
		&=&\frac{2}{1+2|\alpha|^2}\int_0^{\frac1{\sqrt{2}}}\e^{-r^2-|\alpha|^2}r(1-2r^2)I_0(2r|\alpha|)\mathrm{d}r
		,\nonumber\\
		N_{\text{W}_\textrm{odd}}
		&=&\frac{2}{\sqrt{e}}-1 = N_{\text{W}_{\ket{1}}},
\nonumber	}
where $I_0$ is the modified Bessel function. The odd states have a Wigner negative volume equal to that of the single-mode one-photon Fock state, independently of $\alpha$. The situation is different for the even states. When $\alpha=0$, $N_{\text{W}_\textrm{even}}=N_{\text{W}_\textrm{odd}}$ but $N_{\text{W}_\textrm{even}}$ is monotonically decreasing and for $\alpha=1.9$, $N_{\text{W}_\textrm{even}}$ has dropped to 5\% of its maximal value showing that the
 even states  loose their Wigner negativity fast as a function of $\alpha$. 

It is straightforward to compute  the QCS of these states: since they are pure, one can use Eq.~\eqref{eq:pureQCS} directly. One finds
\eq{\Ccal^2_{\psi_{\textrm{even}}}=1+\frac{1}{(1+2|\alpha|^2)^2},\qquad\qquad\Ccal^2_{\psi_{\textrm{odd}}}=2.\nonumber}
Here also, the odd state shows an $\alpha$-independent QCS, which is however lower than the one of the single-mode one-photon Fock state. The QCS of the even states decreases fast with $\alpha$.  
It follows then that, by this criterium also, the odd states are more nonclassical than the even ones but also more prone to environmental decoherence.

One concludes that the odd states have stronger nonclassicality properties than the even ones. The same conclusion can be drawn from the study of the entanglement between the two modes for those states. Indeed, in~\cite{Biagi2020} it is shown to be maximal and independent of $\alpha$ for the odd state.  More precisely, the Negativity of the Partial Transpose (NPT) of these states is
$$
\textrm{NPT}_\textrm{even}=\frac1{1+|\alpha|^2},\qquad \textrm{NPT}_\textrm{odd}= 1,
$$
indicating the odd state  is more strongly entangled than the even one. Again, since the states are pure, one can easily compute their entanglement of formation (EoF) \cite{Bennett}, which has the same behaviour. Since the reduced density matrix is a rank two operator, the maximal possible EoF is $\ln 2$. This is indeed the value reached for the odd states at all values of $\alpha$ as well as  for the even states when $\alpha=0$, as is readily checked. For even states, on the other hand, it tends to its minimal value, which is $0$, as $\alpha$ tends to infinity.

	
\subsection{Photon added two-mode squeezed thermal states (2SqTh+)} \label{s:2modesqth+}
We now briefly consider the case where $\rhoG$ in Eq.~\eqref{eq:twomodep} is a squeezed thermal state. 
We then add one photon with the creation operator $a^\dag(\textbf{c})$ where $\textbf{c}=(c_1\quad \pm\sqrt{1-c_1^2})^T\in \C^2$. The characteristic functions and QCS can be obtained with  Eqs.\eqref{eq:CharacPhotonAddedGaussianState} and \eqref{eq:QCSWch}. Once again,  the formulas are explicit, but not very instructive, and we do not show them here.   It turns out to be  easy to evaluate the Wigner function of the photon-added state at the origin and to observe it does not depend on $\boldsymbol{c}$. The same values are obtained whether the photon is added on the first mode, on the second one, or shared between the two modes. 
One therefore finds, for $q=0.2$, $r=0.5$
$$
N_{\text{W}_{2\text{SqTh}+}}(0.2, 0.5)= 0.104.
$$
which is the same value as for the SqTh$+$ states with the same $r$ and $q$. 

The computation of the QCS and hence of the nonclassicality gain of this state reveals the same phenomenon: they do not depend on $\boldsymbol c$. One finds
$$
\Ccal^2_{2\text{SqTh}+}=1.54=\frac12(\Ccal^2_{\text{SqTh}}+\Ccal^2_{\text{SqTh}+}),
$$
a reflection of the fact that the QCS is the average of the coherence scale of the quadratures.


\section{Conclusion/Discussion}
\label{Conclusion}
We have quantitatively analysed how photon addition/subtraction affects the nonclassical properties of Gaussian states. We concentrated on two measures of nonclassicality, the Wigner negative volume and the quadrature coherence scale (QCS). We have established that, since the QCS tends to undergo a very substantial increase in the photon-addition/subtraction process, the resulting non-Gaussian states are considerably more sensitive to  environmental decoherence than the original Gaussian states. In addition, the decoherence time shortens rapidly with increased squeezing. 

For single-mode fields, we have shown that, whereas at low and intermediate squeezing, photon-addition is considerably more efficient in enhancing or creating nonclassicality than photon subtraction, at high squeezing, the two procedures produce the same effects. The Wigner negative volume saturates in this regime to a noise-dependent fraction of its maximal value, which is attained on photon-added/subtracted squeezed vacuum states. In the course of our analysis, we identified what seems to be a new family of quantum non-Gaussian Wigner positive states, obtained by photon-subtraction from squeezed thermal states.
 
 One may note that the Wigner negativity and the quadrature coherence scale are not the only telltale signs of nonclassicality in quantum optics: 
nonclassicality comes in many guises and can be recognized through the observation of a variety of physical or mathematical properties signaling the quantum nature of the state, the most prominent ones being non-Poissonian statistics, squeezing, Wigner negativity, interference fringes, (quantum) non-Gaussianity and entanglement.
 In the context of quantum optics, a large number of witnesses, measures and monotones of nonclassicality have consequently been developed~\cite{Hillery1,Bach,Hillery2,Hillery3,Lee, Agarwal2,Lee2,Lutkenhaus, Dodonov, Marian,Richter, Kenfack, Asboth, Ryl,Sperling,Killoran,Alexanian,Nair,Ryl2, Yadin, Yadin2, Kwon2,  Takagi18, Horoshko, Luo, Bohmann}. 
It would be  of interest to complete the present study by also testing how these other figures of merit are affected by the photon-addition/subtraction process. Note however that the analytical or even numerical computation of many of those quantities is not straightforward. For example, to compute the Wigner-Yanase skew information one needs a priori to compute the square root of the density matrix, which is not obvious for most states, including the photon-added/subtracted states considered here. Similarly, except for pure states, the quantum Fisher information, that can be used as a nonclassicality measure~\cite{Yadin18, Kwon19}, is generally hard to determine. Let us note that on Gaussian states, the quantum Fisher information coincides with the QCS~\cite{Hertz}, as well as on pure states, but not in general. 

As for the entanglement of multi-mode photon-added/subtracted Gaussian states, it  has been investigated in~\cite{Fabre, FabrePRL17}.  The maximal entanglement increase -- as measured by the Renyi entropy of the Wigner function -- that can occur in the process  has been evaluated in~\cite{Zhang2022}. For pure states, upper bounds on the entanglement of formation in terms of the QCS can be inferred from the results of~\cite{HertzCerfDebievre}.

In experiments, losses are inevitable and any theoretical analysis needs to take them into account in its modelling of the situation. In quantum optics, this is usually done by coupling the field via a beamsplitter to a vacuum mode~\cite{Asboth, se06, Parigi2007, Zavatta, HarocheBook, Biagi2021}, or by simply mixing the state with a vacuum component~\cite{Lvovsky}. In both cases, explicit expressions of the characteristic function of the lossy state are available in terms of the original one. Hence, the methods expounded here can be used to compute both the quadrature coherence scale and Wigner negativity for lossy photon-added/subtracted Gaussian states. They will both be diminished by the losses, as already illustrated in~\cite{Biagi2021} for the Wigner function of single-photon-added thermal states. For the quadrature coherence scale, preliminary computations, not shown here, confirm this picture. Much of the experimental work on those states has concentrated on certifying their nonclassicality~\cite{Zavatta, Parigi2007, Kiesel, Biagi2021}, in particular for high noise and losses, where the quantum nature of the states is strongly suppressed and this certification therefore difficult. Whereas this constitutes an obvious challenge, a perhaps more important challenge is to prepare states of the optical field that show strong nonclassical properties, such as a high value for the Wigner negative volume and/or the squeezing. They can be expected to be more likely to be useful in various quantum technology protocols, but, as we show here, the high value of the quadrature coherence scale generated by the photon-addition/subtraction  makes them strongly sensitive to environmental decoherence and hence hard to prepare and maintain. 

We finally point out that the method to compute the characteristic function of single-photon-added/subtracted states that we introduce here can easily be generalized to the case of multi-photon addition/subtraction and can provide a useful tool for further studies of various features of such states.


\medskip
\noindent {\it Acknowledgments}: 
This work was supported in part by the Agence Nationale de la Recherche under grant ANR-11-LABX-0007-01 (Labex CEMPI) and by the Nord-Pas de Calais Regional Council and the European Regional Development Fund through the Contrat de Projets \'Etat-R\'egion (CPER). 
AH acknowledges financial support from the Natural Sciences and Engineering Research Council of Canada (NSERC). We thank the anonymous referees for their constructive comments that helped us change  the focus of the paper. We are also grateful to G. Patera for helpful discussions. 
\bigskip

\onecolumngrid
%
%


\appendix

\renewcommand{\thesection}{APPENDIX \Alph{section}}

\section{Proof of Eq. \eqref{WignerPlus}}
\label{Appendix B}

We focus on the photon-addition case. Calculations are similiar in the photon-subtraction case. The Wigner function is defined as
\eq{W_+(\boldsymbol{\alpha})=\frac{1}{\pi^{2n}}\int\chi_+(\boldsymbol{ z})e^{(\boldsymbol{\bar z\cdot\alpha}-\boldsymbol{ z\cdot\bar{\alpha}})}d^{2n}\boldsymbol{ z}\nonumber}
where $\chi_+$ is given by Eq.~\eqref{chiPlus}. 
Computing each term of the integral, we have :
\eqarray{\frac{1}{\pi^{2n}}\int (\bar{\boldsymbol{c}}\cdot\partial_{\bar{\boldsymbol z}})(\boldsymbol{c}\cdot\partial_{\boldsymbol z}) \chi(\boldsymbol{ z})e^{\boldsymbol{\bar z\cdot\alpha}-\boldsymbol{ z\cdot\bar{\alpha}}}d^{2n} z&=&\sum_{ij}\bar c_ic_j\frac{1}{\pi^{2n}}\int\left(\partial_{\bar  z_i}\partial_{  z_j}\chi(\boldsymbol{ z})\right)e^{\boldsymbol{\bar z\cdot\alpha}-\boldsymbol{ z\cdot\bar{\alpha}}}d^{2n} z\nonumber\\
	&=&-\sum_{ij}\bar c_ic_j\frac{1}{\pi^{2n}}\int\left(\partial_{  z_j}\chi(\boldsymbol{ z})\right)\alpha_ie^{\boldsymbol{\bar z\cdot\alpha}-\boldsymbol{ z\cdot\bar{\alpha}}}d^{2n} z\nonumber\\
	&=&-\sum_{ij}\bar c_ic_j\alpha_i\bar{\alpha}_j\frac{1}{\pi^{2n}}\int \chi(\boldsymbol{z}) e^{\boldsymbol{\bar z\cdot\alpha}-\boldsymbol{ z\cdot\bar{\alpha}}}d^{2n} z\nonumber\\
	&=&-|\boldsymbol{c}\cdot\bar{\boldsymbol{\alpha}}|^2 W(\boldsymbol{\alpha}\nonumber),
}
where we used an integration by parts. 
Here, $W(\boldsymbol{\alpha})$ is the Wigner function of the initial state. The other terms in the integral are : 
\eqarray{
	\frac{1}{\pi^{2n}}\int (\boldsymbol{c}\cdot\bar{\boldsymbol z})(\bar{\boldsymbol{c}}\cdot\partial_{\bar{\boldsymbol z}}) \chi(\boldsymbol{ z})e^{\boldsymbol{\bar z\cdot\alpha}-\boldsymbol{ z\cdot\bar{\alpha}}}d^{2n} z&=&\sum_{ij} c_i\bar c_j\frac{1}{\pi^{2n}}\int\bar{ z}_i\left(\partial_{\bar z_j}\chi(\boldsymbol{ z})\right)e^{\boldsymbol{\bar z\cdot\alpha}-\boldsymbol{ z\cdot\bar{\alpha}}}d^{2n} z\nonumber\\
	&=&-\sum_{ij} c_i\bar c_j\frac{1}{\pi^{2n}}\left(\int\bar{ z}_i\alpha_j\chi(\boldsymbol{ z})e^{\boldsymbol{\bar z\cdot\alpha}-\boldsymbol{ z\cdot\bar{\alpha}}}d^{2n} z+\delta_{ij}\int\chi(\boldsymbol{ z})e^{\boldsymbol{\bar z\cdot\alpha}-\boldsymbol{ z\cdot\bar{\alpha}}}d^{2n} z\right)\nonumber\\
	&=&-\sum_{ij} c_i\bar c_j\left(\frac{1}{\pi^{2n}}\alpha_j\partial_{\alpha_i}\int\chi(\boldsymbol{ z})e^{\boldsymbol{\bar z\cdot\alpha}-\boldsymbol{ z\cdot\bar{\alpha}}}d^{2n} z+\delta_{ij}W(\boldsymbol{\alpha})\right)\nonumber\\
	&=&-(\bar{\boldsymbol{c}}\cdot \boldsymbol\alpha)(\boldsymbol{c}\cdot\partial_{ \boldsymbol\alpha})W(\boldsymbol{\alpha})-W(\boldsymbol{\alpha}),\nonumber
}
and similarly
\eqarray{
	\frac{1}{\pi^{2n}}\int (\bar{\boldsymbol{c}}\cdot\boldsymbol z)(\boldsymbol{c}\cdot\partial_{\boldsymbol z}) \chi(\boldsymbol{ z})e^{\boldsymbol{\bar z}\cdot\boldsymbol{\alpha}-\boldsymbol{ z}\cdot\boldsymbol{\bar{\alpha}}}d^{2n} z&=&
	-(\boldsymbol{c}\cdot \bar{\boldsymbol\alpha})(\bar{\boldsymbol{c}}\cdot\partial_{ \bar{\boldsymbol\alpha}})W(\boldsymbol{\alpha})-W(\boldsymbol{\alpha}\nonumber),\\
	\frac{1}{\pi^{2n}}\int (\boldsymbol{c}\cdot\bar{\boldsymbol z})(\bar{\boldsymbol{c}}\cdot\boldsymbol z) \chi(\boldsymbol{ z})e^{\boldsymbol{\bar z\cdot\alpha}-\boldsymbol{ z\cdot\bar{\alpha}}}d^{2n}&=&-(\boldsymbol{c}\cdot\partial_{ \boldsymbol\alpha})(\bar{\boldsymbol{c}}\cdot\partial_{ \bar{\boldsymbol\alpha}})W(\boldsymbol{\alpha}).\nonumber
}
Putting everything together, we obtain the Wigner function of the photon added state :
\eqarray{W_+(\boldsymbol{\alpha})&=&
	\mathcal{N}_+
	\left(-\frac12 +|\boldsymbol{c}\cdot\bar{\boldsymbol{\alpha}}|^2 
	-\frac12(\bar{\boldsymbol{c}}\cdot \boldsymbol\alpha)(\boldsymbol{c}\cdot\partial_{ \boldsymbol\alpha})
	-
	\frac12(\boldsymbol{c}\cdot \bar{\boldsymbol\alpha})(\bar{\boldsymbol{c}}\cdot\partial_{ \bar{\boldsymbol\alpha}})
	+\frac14(\boldsymbol{c}\cdot\partial_{ \boldsymbol\alpha})(\bar{\boldsymbol{c}}\cdot\partial_{ \bar{\boldsymbol\alpha}})
	\right)W(\boldsymbol \alpha)\nonumber\\
	&=&\mathcal{N}_+\left[\boldsymbol{c}\cdot\left(\frac{\partial_{ \boldsymbol\alpha}}{2}-\bar{\boldsymbol{\alpha}}\right)\right]\left[\bar{\boldsymbol{c}}\cdot\left(\frac{\partial_{ \bar{\boldsymbol\alpha}}}{2}-\boldsymbol{\alpha}\right)\right]W(\boldsymbol \alpha)
	\nonumber}

\section{Characteristic function of a photon-added/subtracted Gaussian state (proof of Eq. \eqref{eq:CharacPhotonAddedGaussianState})}
\label{Appendix C}

We will compute the characteristic function of a general multi-mode photon-added/subtracted Gaussian state using Eq.~\eqref{chiPlus}. This involves taking derivatives of the Gaussian characteristic function \eqref{chiGaussian} 
\eq{\chiG(\boldsymbol{\xi})=e^{K^{\textrm{G}}(\xi)},\quad \textrm{with}\quad K^{\textrm{G}}(\xi)=-\frac12\boldsymbol{\xi}^T\Omega V \Omega^T\boldsymbol{\xi}-i \sqrt{2}(\Omega \boldsymbol{d})^T\boldsymbol{\xi}\\
\qquad	\text{	and}\qquad \Omega=\bigoplus_{j=1}^n\Omega_2,\qquad\Omega_2=\begin{pmatrix}0&1\\-1&0	\end{pmatrix},\nonumber}
with respect to the complex variables $z_j=\xi_{j1}+i\xi_{j2}\in \C, \ \overline z_j\in \C$. For that purpose, we first recall the expression of $K^{\textrm{G}}$ in terms of these variables. 
We define $\boldsymbol\nu=\frac1{\sqrt{2}}\begin{pmatrix}1&-i\end{pmatrix}^T$ so that $\boldsymbol\xi_\ell=\frac1{\sqrt{2}}(z_\ell\boldsymbol\nu+\bar z_\ell\bar{\boldsymbol\nu})$. Since $\Omega_2\boldsymbol \nu=-i\boldsymbol\nu$ and $\Omega_2\bar{\boldsymbol\nu}=i\bar{\boldsymbol\nu}$, we find
$$
\Omega_2\boldsymbol\xi_\ell = i\frac1{\sqrt2} u^T\Omega_2 \begin{pmatrix} z_\ell \\ \overline z_\ell\end{pmatrix},\quad{\textrm{where}}\quad u=\begin{pmatrix} \bar {\boldsymbol\nu}^T\\\boldsymbol{\nu}^T\end{pmatrix}=\frac1{\sqrt{2}}\begin{pmatrix}1&i\\1&-i\end{pmatrix}.
$$
  Introducing the unitary matrix
 $
  U=\bigoplus_ju,
 $
 and defining $A=UVU^T$, we find that $A$ is the matrix of the covariances of the creation and annihilation operators:
	\eq{\label{eq:Adef}
	A=\begin{pmatrix}
		\tilde A^{11}&\tilde A^{12} &\cdots&\tilde A^{1n}\\\tilde A^{21}& & &\vdots\\\vdots& & &\\\tilde A^{n1}&\cdots& &\tilde A^{nn}
	\end{pmatrix}\qquad\text{with}\qquad\tilde A^{ij}=2\begin{pmatrix}\cov{a_i,a_j}&\cov{a_i,a^\dag_j}\\\cov{a^\dag_i,a_j}&\cov{a^\dag_i,a^\dag_j}.
\end{pmatrix}=u\tilde{V}^{ij}u^T\nonumber.}
Here $\tilde V^{ij}$ is the two-by-two submatrix of the covariance matrix $V$ defined by
\begin{equation*}\label{eq:tildeV}
\tilde{V}^{ij} =\begin{pmatrix} V_{2i-1,2j-1}&V_{2i-1,2j}\\V_{2i,2j-1}&V_{2i,2j}
\end{pmatrix}
=2\begin{pmatrix} \cov{\hat x_i, \hat x_j}& \cov{\hat x_i, \hat p_j}\\
\cov{\hat p_i, \hat x_j}&\cov{\hat p_i, \hat p_j}\end{pmatrix}.
\end{equation*}
One then finds
\eqarray{\frac12\boldsymbol \xi^T\Omega V\Omega^T\boldsymbol\xi&=&\frac12\sum_{kl}\boldsymbol\xi_k^T(\Omega_2\tilde{V}^{kl}\Omega_2)\boldsymbol\xi_l
=-\frac14\sum_{kl}\begin{pmatrix} z_k &\bar z_k\end{pmatrix} \Omega_2^T\tilde A^{kl}\Omega_2 \begin{pmatrix}z_l \\\bar z_l\end{pmatrix}\nonumber
}
and
\eqarray{i\sqrt{2}(\Omega \boldsymbol d)^T\boldsymbol\xi=  \sum_k \boldsymbol d_k^Tu^T\Omega_2\begin{pmatrix} z_k\\ \bar z_k\end{pmatrix}.\nonumber} 
Using that
$$
\boldsymbol  d_k^Tu^T=\boldsymbol  d_k^T(\bar{\boldsymbol \nu}\  \boldsymbol \nu)=\begin{pmatrix} \mean{a_k}&\mean{a_k^\dag} \end{pmatrix},
$$
this leads to
$$
K^{\textrm{G}}(\boldsymbol{Z})=\frac14\boldsymbol{Z}^T(\Omega^TA\Omega)\boldsymbol{Z}-\Delta^T\Omega \boldsymbol{Z},
$$
where
$$
\boldsymbol{Z}=\begin{pmatrix} z_1\\ \overline z_1\\\vdots \\ z_n\\ \overline{z}_n \end{pmatrix},\qquad \textrm{and}\qquad 
\Delta=\begin{pmatrix}\mean{a_1}\\\mean{a^\dagger_1}\\\vdots \\ \mean{a_n}\\\mean{a^\dagger_n}\end{pmatrix}=U\boldsymbol{ d}.
$$
It is now easy to take the derivatives along $z_k$ and $\bar z_k$ and we obtain:
\eqarray{
\boldsymbol	c\cdot \partial_{ z}\chiG(\boldsymbol z)&=&\left(\cov{a^\dag(\boldsymbol c),a^\dag(\boldsymbol z)-a(\boldsymbol{z})}+\mean{a^\dag(\boldsymbol c)}	\right)\chiG(\boldsymbol z),\nonumber\\
	\overline{\boldsymbol c}\cdot\partial_{ \bar z}\chiG(\boldsymbol z)&=&-\Big(\cov{a(\boldsymbol c),a^\dag(\boldsymbol z)-a(\boldsymbol{z})}+\mean{a(\boldsymbol c)}	\Big)\chiG(\boldsymbol z),\nonumber\\
		(\overline{\boldsymbol c} \cdot \partial_{\bar z})( \boldsymbol c\cdot\partial_{ z})\chiG(\boldsymbol z)&=&-\Big(\cov{a^\dag(\boldsymbol c) ,a^\dag(\boldsymbol z)-a(\boldsymbol{z})}+\mean{a^\dag(\boldsymbol c)}	\Big)
	\Big(\cov{a(\boldsymbol c),a^\dag(\boldsymbol z)-a(\boldsymbol{z})}+\mean{a(\boldsymbol c)}	\Big)\chiG(\boldsymbol z)\nonumber\\
	&&\qquad\qquad\qquad\hskip 6cm -\cov{a^\dag(\boldsymbol c),a(\boldsymbol c)}\chiG(\boldsymbol z).
	\nonumber
}
According to Eq. \eqref{chiPlus}, the characteristic function  of the photon added/subtracted state is given by 
 \eqarray{\chiGpm(\boldsymbol z)&=&\mathcal{N}_\pm\Bigg(\pm\frac{\chiG(\boldsymbol z)}{2}-(\overline{\boldsymbol c}\cdot\partial_{\bar z})
 	(\boldsymbol c\cdot\partial_{ z})\chiG(\boldsymbol z)\pm\frac{\boldsymbol c\cdot \bar z}{2}(\overline{\boldsymbol c}\cdot \partial_{\bar z})\chiG(\boldsymbol z)\pm\frac{\overline{\boldsymbol c}\cdot z}{2}(\boldsymbol c\cdot\partial_{z})\chiG(\boldsymbol z)-\frac{|\overline{\boldsymbol c}\cdot z|^2}{4}\chiG(\boldsymbol z)	\Bigg).\nonumber
	}
Hence we obtain
	 \eq{\chiGpm(\boldsymbol z)
	=\mathcal{N}_\pm \left( \text{Cov}[a^\dag(\boldsymbol{c}),a(\boldsymbol{c})] \pm\frac12-\Big(\boldsymbol{ c}\cdot\boldsymbol{\gamma}_\pm\Big)\Big(
	\bar{\boldsymbol{ c}}\cdot\boldsymbol{\delta}_\pm\Big)
	\right)\chiG(\boldsymbol z),\nonumber}
with
$$(\gamma_\pm)_k=\text{Cov}[a^\dag_k,a^\dag(\boldsymbol{z})-a(\boldsymbol{z})]\mp \frac12\bar{z}_k+\mean{a^\dag_k}\qquad\qquad
(\delta_\pm)_k=-\text{Cov}[a_k,a^\dag(\boldsymbol{z})-a(\boldsymbol{z})]\mp\frac12 {z_k}-\mean{a_k}.$$
This expression can be further simplified as follows.  Note that $\overline U\  U^\dag=\bigoplus_{j=1}^n \sigma_x$ with $\sigma_x=\begin{pmatrix} 0&1\\ 1&0\end{pmatrix}$ and $
\boldsymbol{ d}^T=~\begin{pmatrix}
\mean{a_1}&\mean{a_1^\dag}&\cdots&\mean{a_n}& \mean{a_n^\dag}\end{pmatrix}\overline U$. 
Using $U^T\Omega U=-i\Omega$ and the unitarity of $U$, one then finds
\begin{eqnarray*}
\begin{pmatrix} \gamma_1& \delta_1&\dots&\gamma_n&\delta_n\end{pmatrix}^T_\pm&=&\frac12\left(\Omega^TA\Omega \boldsymbol{Z}\mp\overline U U^\dag \boldsymbol{Z}\right)-\Omega U \boldsymbol{ d}\\
&=&\overline U\left(\frac12\left(\Omega V\Omega\mp \bbone\right) U^\dag \boldsymbol{Z}+i\Omega \boldsymbol{d}\right).
\end{eqnarray*}
Recalling from Eq.~\eqref{eq:mc} that, 
 for all $c\in\C^n$, 
$$
\boldsymbol{m}_{\boldsymbol{c}}=\bar U^T\begin{pmatrix}c_1\\0\\ \vdots \\  c_n\\ 0 \end{pmatrix}=\frac{1}{\sqrt 2}\begin{pmatrix} c_1\\- i c_1\\ \vdots\\ c_n\\- i  c_n\end{pmatrix}\in\C^{2n},
$$
we have
$$
\overline{\boldsymbol{m}}_{\boldsymbol{c}}^TU^T=\begin{pmatrix} 0&\bar c_1&\dots&0&\bar c_n\end{pmatrix}, \quad \text{Cov}[a^\dag(\boldsymbol{c}),a(\boldsymbol{c})]=\frac12\overline{\boldsymbol{m}}_{\boldsymbol{c}}^TV\boldsymbol{m}_{\boldsymbol{c}}.
$$
The term $(\boldsymbol c\cdot\boldsymbol\gamma_\pm)(\bar{\boldsymbol c}\cdot\boldsymbol \delta_\pm)$ can thus be written as
\eqarray{(\boldsymbol c\cdot\boldsymbol\gamma_\pm)(\bar{\boldsymbol c}\cdot\boldsymbol \delta_\pm)&=&\begin{pmatrix}
		\gamma_1&\delta_1&	\gamma_2&\delta_2&\hdots&	\gamma_n&\delta_n
	\end{pmatrix}_\pm      \begin{pmatrix}c_1\\0\\ \vdots \\  c_n\\ 0 \end{pmatrix}
	\begin{pmatrix} 0&\bar c_1&\dots&0&\bar c_n\end{pmatrix}  
	\begin{pmatrix}
		\gamma_1\\\delta_1\\	\gamma_2\\\delta_2\\\vdots\\	\gamma_n\\\delta_n
	\end{pmatrix}_\pm\nonumber\\
	&=&\left[\boldsymbol{ Z}^T\bar U\frac12\left(\Omega V\Omega\mp\bbone\right)+i\boldsymbol{ d}^T\Omega^T\right] 
	\boldsymbol{m}_{\boldsymbol{c}} \overline{\boldsymbol{m}}_{\boldsymbol{c}}^T
	\left[\frac12\left(\Omega V \Omega\mp\bbone\right)\bar U^T\boldsymbol{ Z}+i\Omega\boldsymbol{ d}\right].\nonumber
}
 
The characteristic function can thus be written 
	\eqarray{\chiGpm(\boldsymbol z)
		=\mathcal{N}_\pm \left( \frac12 
		\overline{\boldsymbol{m}}_{\boldsymbol{c}}^TV\boldsymbol{m}_{\boldsymbol{c}}\pm\frac12
		-\boldsymbol{ \beta}^T_\pm \boldsymbol{m}_{\boldsymbol{c}}\overline{\boldsymbol{m}}_{\boldsymbol{c}}^T\boldsymbol{ \beta}_\pm
		\right)\chiG(\boldsymbol z)\nonumber}
	with $\boldsymbol{ \beta}_\pm=\frac12\left(\Omega V \Omega\mp\bbone\right) U^\dag\boldsymbol{ Z}+i\Omega\boldsymbol{ d}.$ This is Eq. \eqref{eq:CharacPhotonAddedGaussianState} .




\section{Wigner function of a photon added/subtracted state (proof of Eq.\eqref{eq:Wignerpm})}
\label{Appendix D}
To derive the expression in Eq.~\eqref{eq:Wignerpm}, we proceed similarly. Note first that, using Eq.~\eqref{FourierWigner}, one readily computes the well known Wigner function of a Gaussian state with characteristic function $\chi^{\textrm{G}}$. It reads
\eq{\WG(\boldsymbol\alpha)=
	\frac{2^n}{\pi^n\sqrt{\det V}}\exp
\left\{-Y^T
A^{-1}Y\right\}\nonumber
}
where $Y=(\alpha_1-\mean{a_1},\bar \alpha_1-\mean{a_1^\dag}, \hdots,\alpha_n-\mean{a_n}, \bar \alpha_n-\mean{a_n^\dag})\in\C^{2n}$. 
One then readily computes the $\alpha_k$ and $\overline \alpha_k$ derivatives of $\WG(\boldsymbol{ \alpha})$. Inserting them in Eq. \eqref{WignerPlus}, using the definition of $\boldsymbol{m}_{\boldsymbol{c}}$ in Eq.~\eqref{eq:mc} and
\begin{equation*}\label{eq:Ainv}
A^{-1}=\overline U V^{-1} U^\dag,
\end{equation*}
 one obtains the Wigner function of the photon-added/subtracted state:
\eqstar{\WGpm(\boldsymbol\alpha)
	=\mathcal{N}_\pm\Bigg((\boldsymbol c\cdot\boldsymbol\mu_\pm)(\bar{\boldsymbol c}\cdot\boldsymbol \eta_\pm)+M_\pm(V,\boldsymbol{c})\Bigg)\WG(\boldsymbol\alpha)
}
where $M_\pm(V,\boldsymbol{c})${$\in\R$} is independent of $\boldsymbol{\alpha}$ and given by
\begin{equation*}
M_\pm(V,\boldsymbol{c})=\mp\frac12-\frac12\begin{pmatrix} c_1&0&\dots&c_n&0\end{pmatrix} A^{-1}\begin{pmatrix}0\\\bar c_1\\ \vdots \\ 0 \\ \bar c_n\end{pmatrix}
=\mp\frac12-\frac12\overline{\boldsymbol{m}}_{\boldsymbol{c}}^T V^{-1} \boldsymbol{m}_{\boldsymbol{c}},
\end{equation*}
and where the vectors $\mu_\pm,\eta_\pm\in\C^n$ are defined by
\eqstar{\begin{pmatrix}
		\mu_1\\\eta_1\\	\mu_2\\\eta_2\\\vdots\\	\mu_n\\\eta_n
	\end{pmatrix}_\pm=\overline U\left(V^{-1}\pm\bbone\right) U^\dag \begin{pmatrix}
		\alpha_1\\\bar \alpha_1\\\alpha_2\\\bar \alpha_2\\\vdots\\\alpha_n\\\bar \alpha_n
	\end{pmatrix}-A^{-1} \begin{pmatrix}
		\mean{a_1}\\\mean{a_1^\dag}\\\mean{a_2}\\\mean{a_2^\dag}\\\vdots\\\mean{a_n}\\\mean{a_n^\dag}
	\end{pmatrix}=\overline U\left(V^{-1}\pm\bbone\right) \boldsymbol{ r}-\overline UV^{-1} \boldsymbol{ d}.
}
Here we used the fact that the vector of quadratures $\boldsymbol{ r}\in \R^{2n}$ and the vector of displacement $\boldsymbol d\in\R{^2}$ can be written as $\boldsymbol{ r}^T=\begin{pmatrix}
\alpha_1&\bar \alpha_1&\cdots&\alpha_n&\bar \alpha_n\end{pmatrix}\overline U $ and $
\boldsymbol{ d}^T=~\begin{pmatrix}
\mean{a_1}&\mean{a_1^\dag}&\cdots&\mean{a_n}& \mean{a_n^\dag}\end{pmatrix}\overline U$.
Using $U^T\Omega U=-i\Omega$, the term $(\boldsymbol c\cdot\boldsymbol\mu_\pm)(\bar{\boldsymbol c}\cdot\boldsymbol \eta_\pm)$ can be rewritten as follows:
\eqarray{(\boldsymbol c\cdot\boldsymbol\mu_\pm)(\bar{\boldsymbol c}\cdot\boldsymbol \eta_\pm)&=&\begin{pmatrix}
		\mu_1&\eta_1&	\mu_2&\eta_2&\hdots&	\mu_n&\eta_n
	\end{pmatrix}_\pm      \begin{pmatrix}c_1\\0\\ \vdots \\  c_n\\ 0 \end{pmatrix}
	\begin{pmatrix} 0&\bar c_1&\dots&0&\bar c_n\end{pmatrix}  
	\begin{pmatrix}
		\mu_1\\\eta_1\\	\mu_2\\\eta_2\\\vdots\\	\mu_n\\\eta_n
	\end{pmatrix}_\pm\nonumber\\
	&=&\left[\boldsymbol{ r}^T\left(V^{-1}\pm\bbone\right)-\boldsymbol{ d}^TV^{-1}\right]  U^\dag\begin{pmatrix}c_1\\0\\ \vdots \\  c_n\\ 0 \end{pmatrix}
	\begin{pmatrix} 0&\bar c_1&\dots&0&\bar c_n\end{pmatrix}  \overline U\left[\left(V^{-1}\pm\bbone\right)\boldsymbol{ r}-V^{-1}\boldsymbol{ d}\right]\nonumber\\
	&=&\left[\boldsymbol{ r}^T\left(V^{-1}\pm\bbone\right)-\boldsymbol{ d}^TV^{-1}\right] 
	\boldsymbol{m}_{\boldsymbol{c}} \overline{\boldsymbol{m}}_{\boldsymbol{c}}^T
	\left[\left(V^{-1}\pm\bbone\right)\boldsymbol{ r}-V^{-1}\boldsymbol{ d}\right].\nonumber
}
Introducing
\begin{equation*}\label{eq:lambda}
\boldsymbol{\lambda}_\pm=\left(V^{-1}\pm\bbone\right)\boldsymbol{ r}-V^{-1}\boldsymbol{ d}\in\R^{2n},
\end{equation*}
this yields
\eqstar{\WGpm(\boldsymbol r)
	=\mathcal{N}_\pm\Bigg(M_\pm(V,\boldsymbol{c})+	\boldsymbol{ \lambda}_\pm^T
	\boldsymbol{m}_{\boldsymbol{c}} \overline{\boldsymbol{m}}_{\boldsymbol{c}}^T
	\boldsymbol{ \lambda}_\pm
\Bigg)\WG(\boldsymbol r)
}
which is Eq.~\eqref{eq:Wignerpm}. 
%


\section{QCS of the $\text{SqTh}+$ and $\text{SqTh}-$ states}
\label{Appendix D bis}

With the characteristic function \eqref{ChiSqThPlus}  and Eq. \eqref{eq:QCSWch} we find the value of the QCS  of the $\text{SqTh}+$ state :
\eqarray{\Ccal^2_{\text{SqTh}+}(q,r)&=&\frac{(1-q)/(1+q)}{ 2 \left(1-q^4\right) \cosh 2 r+2\left(1+q^2\right)^2+\left(q^4+10 q^2+1\right) \sinh ^22 r}\nonumber\\
	&&	\times\left[
	-8 q \left(q^2-1\right)+3 \left(q^4-4 q^3+10 q^2-4 q+1\right) \cosh ^32 r+6 (q-1)^2 \left(1-q^2\right) \cosh ^22 r\right.\nonumber\\
	&&\left.+\left(3 q^4+8 q^3-26 q^2+8 q+3\right) \cosh 2 r\right]\nonumber}

One then readily computes
\eq{\lim_{r\to+\infty}\mathcal{R}_{\text{SqTh}+}(q,r)=2-12q\frac{q^2+1}{q^4+10q^2+1}.\nonumber}

Similarly, with the characteristic function
\eqarray{\chi_{\text{SqTh}-}(\boldsymbol z)&=&\chi_{\text{SqTh}}(\boldsymbol z)\left(\frac{2 q |\boldsymbol  z|^2}{\left(1-q^2\right) \cosh 2 r-(1-q)^2}+\frac{q+1}{q-1} \left(e^{2 r}  \xi_1^2+e^{-2 r}  \xi_2^2\right)+1\right).\nonumber}
  and Eq. \eqref{eq:QCSWch} we find the value of the QCS  of the $\text{SqTh}-$ state :
\eqarray{\Ccal^2_{\text{SqTh}-}(q,r)&=&
	\frac{(1-q) \sqrt{\frac{1}{q+1}} }{2 \sqrt{q+1} \left(4 \left(q^4-1\right) \cosh (2 r)+3 q^4-2 q^2+\left(q^4+10 q^2+1\right) \cosh (4 r)+3\right)}\nonumber\\
	&&\times \left[12 (q+1) (q-1)^3 \cosh (4 r)+(21 q^4-4 q^3-14 q^2-4 q+21) \cosh (2 r)\right.\nonumber\\
	&&\left.+3 (1 - 4 q + 10 q^2 - 4 q^3 + q^4) \cosh (6 r)+4 (q+1) (3 q^2+2q+3) (q-1)\right]\nonumber
}
and
\eq{\lim_{r\to+\infty}\mathcal{R}_{\text{SqTh}-}(q,r)=2-12q\frac{ \left(q^2+1\right)}{q^4+10 q^2+1}=\lim_{r\to+\infty}\mathcal{R}_{\text{SqTh}+}(q,r).\nonumber}

\section{Wigner negative volume of the $\text{SqTh}+$ and $\text{SqTh}-$ states}
\label{Appendix E}
The Wigner negative volume~\cite{Kenfack}, denoted by $N_{\text{W}}(\rho)$ is defined as  the absolute value of the integral of the Wigner function over the area where the latter is negative. 
The Wigner function of the $\text{SqTh}+$ state is computed with \eqref{eq:Wignerpm}  and  we obtain 
\eqarray{W_{\text{SqTh}+}(x,p)&=&\frac{2(1-q)^2 }{\pi\left(  (1+q)^2 \cosh (2 r)+1-q^2\right)} \exp \left(-\frac{(1-q) \left(e^{2 r}x^2 +e^{-2 r} p^2\right)}{1+q}\right)\nonumber\\
	&&\times \left(\left( \frac{ 1-q}{1+q}e^{2 r}+1\right)^2 x^2 + \left( \frac{1-q}{1+q}e^{-2 r}+1\right)^2  p^2 -\frac{1-q}{1+q} \cosh (2r) -1\right) \nonumber}
We easily see that 
the Wigner function of a  $\text{SqTh}+$ state is negative inside the ellipse 
\eq{\left( \frac{1-q }{1+q}e^{2 r}+1\right)^2x^2+ \left(\frac{1-q }{1+q}e^{-2 r}+1\right)^2p^2 =1+\frac{ (1-q) \cosh (2 r)}{1+q}.\nonumber}
The  semi-major and semi-minor axes are given by
\eq{\kappa_x=\frac{e^{-r} \sqrt{(1+q)^2+\left(1-q^2\right) \cosh (2 r)}}{ 2(\cosh r-q \sinh r)}\qquad
	\kappa_p=\frac{e^r \sqrt{(1+q)^2 +(1-q^2) \cosh (2 r)}}{ 2 (\cosh r+q \sinh r)}\nonumber} 
and the Wigner function reaches its minimal value at the origin:
\eq{W_{\text{SqTh}+}(0)=-\frac{ (1-q)^2 ((1-q) \cosh (2 r)+1+q)}{\pi  (1+q)^2 ((1+q) \cosh (2 r)+1-q)}.\nonumber}
At large squeezing, the Wigner negative volume of these states saturates to a value that decreases with increasing temperature $q$. To see this, we note that, at large $r$, one has, with $\tilde x=\e^r x, \tilde p=\e^{-r}p$ and $\mu=\frac{1-q}{1+q}$, that
$$
W_{\text{SqTh}+}(x,p)\approx \frac{4}{\pi}\mu^3 \e^{-\mu(\tilde x^2+\tilde p^2)}\left(\mu \tilde x^2 +\mu^{-1}\tilde p^2-\frac12\right).
$$
It then follows from a straightforward computation that
$$
N_{W, \text{SqTh}+}(q, +\infty):=N_{\text{W}}(\rho_{\text{SqTh}+})(q,+\infty)=\frac{\mu^3}{\pi}\left|\int_0^{2\pi}\left[\frac{2-a(\mu,\theta)}{2a(\mu,\theta)}-a(\mu,\theta)^{-1}\e^{-\frac12a(\mu,\theta)}\right]\rd \theta\right|
$$
where
$$
a(\mu,\theta)=\cos^2\theta +\mu^2\sin^2\theta.
$$
When $q=0$, one has $\mu=1$ and $a(\mu,\theta)=1$ and hence
$$
N_{\text{W}}(\rho_{\text{SqTh}+})(0,+\infty)=\frac{2}{\sqrt \mathrm{e}}-1,
$$
as expected in view of Eq.~\eqref{eq:NWSqVp}. As a result, for small $q$ one has approximately
$$
N_{\text{W}, \text{SqTh}+}(q,+\infty)\simeq \left(\frac{2}{\sqrt \mathrm{e}}-1\right) \mu^3.
$$

	Similarly, the Wigner function of the $\text{SqTh}-$ state is computed with \eqref{eq:Wignerpm}  and  we obtain
	\eqarray{W_{\text{SqTh}-}(x,p)&=&\frac{2(1-q)^2 }{\pi\left(  (1+q)^2 \cosh (2 r)-1+q^2\right)} \exp \left(-\frac{(1-q) \left(e^{2 r}x^2 +e^{-2 r} p^2\right)}{1+q}\right)\nonumber\\
		&&\times \left(\left( 1-\frac{ 1-q}{1+q}e^{2 r}\right)^2 x^2 + \left( 1 -\frac{1-q}{1+q}e^{-2 r}\right)^2  p^2 -\frac{1-q}{1+q} \cosh (2r) +1\right) \nonumber}
	We easily see that 
	the Wigner function of a  $\text{SqTh}-$ state is negative inside the ellipse 
	\eq{\left(1- \frac{1-q }{1+q}e^{2 r}\right)^2x^2+ \left(1-\frac{1-q }{1+q}e^{-2 r}\right)^2p^2 =-1+\frac{ (1-q) \cosh (2 r)}{1+q}.\nonumber}
	provided that $q<\tanh ^2(r)$. Otherwise, the Wigner function is always positive. Remark that when $q,r=0$, we get $\mathcal{N}_-=\infty$ and the Wigner function is not defined.
	
	The  semi-major and semi-minor axes are given by
	\eq{\kappa_x=\frac{e^{-r} \sqrt{\left(1-q^2\right) \cosh (2 r)-(1+q)^2}}{2(\sinh (r)-q \cosh (r))}\qquad
		\kappa_p=\frac{e^r \sqrt{\left(1-q^2\right) \cosh (2 r)-(1+q)^2}}{2 (\sinh (r)+q \cosh (r))}\nonumber\\
	} 	and the Wigner function reaches its minimal value at the origin:
	\eq{W_{\text{SqTh}-}(0)=\frac{ (1-q)^2 (-(1-q) \cosh (2 r)+1+q)}{\pi  (1+q)^2 ((1+q) \cosh (2 r)-1+q)}.\nonumber}

It is readily checked that the asymptotic behaviour of $W_{\text{SqTh}-}(x,p)$ is, for large $r$,  identical to that of $W_{\text{SqTh}+}(x,p)$.

\vfill
\eject

\twocolumngrid


\end{document}